\RequirePackage{silence}
\documentclass[]{aa}
\usepackage{amsmath}
\usepackage[dvipsnames]{xcolor}
\usepackage{graphicx}
\usepackage{hyperref}
\usepackage{multirow}
\usepackage{amsmath}
\usepackage[normalem]{ulem}
\usepackage[nameinlink]{cleveref}
\usepackage[multiple]{footmisc}
\crefname{paragraph}{\S}{\S\S} 
\newcommand{\dd}{\mathrm{d}}

\providecommand{\bjdtdb}{\ensuremath{\rm {BJD_{TDB}}}}
\providecommand{\tjdtdb}{\ensuremath{\rm {TJD_{TDB}}}}

\providecommand{\msun}{\ensuremath{M_\odot}}
\providecommand{\rsun}{\ensuremath{\,R_\odot}}
\providecommand{\lsun}{\ensuremath{\,L_\odot}}
\providecommand{\mj}{\ensuremath{\,M_{\rm J}}}
\providecommand{\rj}{\ensuremath{\,R_{\rm J}}}

\providecommand{\fave}{\langle F \rangle}
\providecommand{\fluxcgs}{10$^9$ erg s$^{-1}$ cm$^{-2}$}

\usepackage{natbib}
\usepackage{scalerel}
\usepackage{comment}
\usepackage[table]{xcolor}
\newcommand{\exofast}{\texttt{EXOFASTv2}\xspace}

\begin{document}
\title{
The high-eccentricity exoplanet HD 80606 b: Probing \\general relativistic apsidal precession and tidal effects}

\author{Siska-Lut Buelens\inst{1,2} \and Erica Thygesen\inst{1} \and Neda Heidari\inst{2} \and Georgia S. C. Stewart\inst{2} \and \\ Luc Blanchet\inst{2}\thanks{luc.blanchet@cnrs.fr} \and Jean-Philippe Beaulieu\inst{1,2}\thanks{jean-philippe.beaulieu@cnrs.fr}}
\institute{
 Greenhill Observatory  and School of Natural Sciences, University of Tasmania, P.O. Box 807, Sandy Bay, TAS 7006 Australia
 \and
  Institut d'Astrophysique de Paris,
UMR 7095, CNRS, Sorbonne Universit\'e, 98 bis bd Arago, 75014 Paris, France
}

\titlerunning{Probing
general relativistic apsidal precession and tidal effects}
   \authorrunning{Buelens et al.}

\date{To be submitted  to A\&A}

\abstract
{ HD 80606 b is a highly eccentric giant exoplanet ($e \simeq 0.93$)
exhibiting both transits and secondary eclipses. The availability of these observations over a long temporal baseline
makes it a favourable system for detecting relativistic apsidal
precession through precision timing measurements.}
{ We search for apsidal precession through changes in
transit--eclipse timing separations and assess the
contributions of general relativity, tidal deformation,
and possible unseen planetary companions.}
{We combine Spitzer, TESS, and JWST photometry with more than two decades
of radial-velocity measurements to derive updated stellar and orbital
parameters using \exofast and a Keplerian orbital model.
We individually fit transit and secondary-eclipse light curves to
measure their midpoints, then apply the formalism of \citet{blanchet2019detecting} to construct a timing observable sensitive to
secular changes in the transit--eclipse separation.
We compare measured shifts with predictions incorporating general
relativity and Newtonian tidal and rotational deformations.}
{Comparing the transit–eclipse timing separations derived from the earlier Spitzer observations and the later TESS–MIRI observations, we measure a shift of
$507.9 \pm 157.0$\,s relative to the zero-precession prediction,
corresponding to a $3.24\sigma$ deviation.
The observed shift exceeds the general-relativistic prediction of
$187.67$\,s by $2.04\sigma$.
Including tidal deformation reduces this discrepancy to below
$2\sigma$, although the measured value remains above the nominal
predictions.

}
{We find evidence for apsidal precession in HD 80606 b, providing a first indication of such orbital evolution in an exoplanetary system where general relativity is expected to dominate. The observed timing shift is consistent within $2\sigma$ with the combined contributions of general relativity and tidal deformation, although the relativistic contribution cannot yet be isolated independently. Future observations with JWST and Ariel could further constrain these effects and investigate possible additional dynamical perturbations.
}

 \maketitle
 \nolinenumbers

\section{Introduction}

Recent advances in high-precision transit and secondary eclipse observations have opened the possibility of testing subtle deviations from strictly Keplerian orbital motion in selected exoplanet systems \citep{millholland2025exploring,jordan2008observability}. One manifestation of such a deviation is apsidal precession: the gradual rotation of the line of apsides (equivalently a gradual change in the argument of periastron, $\omega$), causing successive orbits to differ slightly in orientation \citep{antoniciello2021detecting, jordan2008observability}. Apsidal precession may arise from several phenomena, including general relativistic effects, tidal and rotational   distortions of the planet and host star, and gravitational perturbations from additional companion planets \citep{blanchet2019detecting}. 

The change in orbital orientation causes the predicted times of primary transit and secondary eclipse to shift accordingly. These timing offsets are particularly sensitive in eccentric systems, where the interval between transit and eclipse depends strongly on the argument of periastron \citep{barnes2007effects}. Therefore, highly eccentric exoplanets provide an opportunity to test whether small non-Keplerian effects can be measured directly through precision timing observations.

Observational evidence for apsidal precession has been reported
in several exoplanetary systems, including WASP-19 b and
WASP-43 b \citep{Bernabo2024,Bernabo2025}, while apsidal
precession has also been proposed as an explanation for
timing variations in KELT-9 b \citep{Harre2023}.
However, in these systems the general-relativistic contribution
is relatively small compared with the observed precession,
which is dominated by classical effects or remains
incompletely explained. 
A general-relativistic contribution to exoplanetary apsidal precession has therefore yet to be unambiguously established observationally.

HD 80606 b represents a particularly favourable system for searching for observational signatures of apsidal precession, as general relativity (GR) is expected to dominate its apsidal evolution \citep{blanchet2019detecting}. Discovered through radial-velocity measurements by \citet{Naef2001HD80606b}, the planet occupies an unusually eccentric orbit ($e \simeq 0.93$) with a period of approximately 111 days. HD 80606 b is also one of the few highly eccentric giant exoplanets for which both transits and secondary eclipses have been observed, beginning with the Spitzer eclipse detection by \citet{Laughlin2009HD80606b} and subsequent transit observations constraining the orbital configuration of the system \citep{Moutou2009HD80606b, Fossey2009HD80606b, Winn2009HD80606b, Hebrard2010HD80606b}. Using these properties, \citet{blanchet2019detecting} investigated whether apsidal precession in HD 80606 b could become observable in the near future. They calculated a general relativistic contribution to the apsidal precession rate of approximately $215$ arcsec century$^{-1}$, with an additional tidal contribution of approximately $34$ arcsec century$^{-1}$. Rather than attempting to directly measure changes in $\omega$, they showed that the cumulative effect should manifest observationally as a drift in the relative timing of transits and secondary eclipses, corresponding to an expected shift of order several minutes over a decade-long baseline. This raises the question of whether the observational baseline now available is sufficient to reveal the predicted apsidal evolution. 

More recently, \citet{beichman2026eclipse} investigated possible orbital evolution in HD 80606 b using radial-velocity measurements together with published transit and secondary-eclipse timings, including recent JWST observations. They found that a model with constant period, eccentricity, and argument of periastron adequately describes the available data, while allowing a linear apsidal precession rate yielded $\dot{\omega} = 302^{+294}_{-289}$ arcsec century$^{-1}$, statistically consistent with both zero and the expected general-relativistic contribution. Their analysis therefore found no significant evidence for apsidal precession with the currently available timing data.

Since the analysis of \citet{blanchet2019detecting}, additional transit and secondary-eclipse observations have become available, enabling tests of the predicted long-term timing evolution. In the present work, we combine archival radial-velocity measurements compiled by \citet{pearson2022utilizing} with transit observations from \textit{Spitzer} and \textit{TESS} \citep{Hebrard2010HD80606b, pearson2022utilizing}, together with more recent secondary-eclipse observations obtained using JWST. These include published NIRSpec observations \citep{Sikora2025HD80606b} as well as publicly available MIRI observations analysed in this work.

Our analysis follows an approach complementary to that of \citet{beichman2026eclipse}. Rather than fitting a secular precession rate directly to the measured event times and radial-velocity observations, we perform a new global fit to the underlying photometric and radial-velocity datasets using a purely Newtonian (Keplerian) orbital model. This provides the reference orbital solution required by the perturbative formalism of \citet{blanchet2019detecting}, in which relativistic and tidal corrections are calculated relative to the Newtonian orbit.

 The timing datasets also differ: in particular, our analysis includes the 2010 Spitzer transit used as the reference epoch in the formalism of \citet{blanchet2019detecting}, whereas \citet{beichman2026eclipse}  include additional transit and eclipse epochs not considered here. Using the revised system parameters, we independently measure individual transit and secondary-eclipse midpoints and apply the formalism of \citet{blanchet2019detecting} to test the predicted secular evolution of a specific transit–eclipse timing combination. This approach tests the departure from the zero-precession prediction through a dedicated timing observable, rather than through a global comparison of precessing and non-precessing orbital models. The two analyses therefore provide complementary, but not directly equivalent, statistical tests of apsidal precession.

The remainder of this paper is organised as follows. Section \ref{sec: data} describes the photometric and radial-velocity observations used in the analysis. Section \ref{sec:updated_system_params} presents the updated determination of the system parameters. Section \ref{sec:theoretical predictions} revisits the theoretical predictions for apsidal precession using the revised orbital solution. Section \ref{sec: timing measurements} presents the individual transit and eclipse timing measurements and compares their evolution with the theoretical expectations. Section \ref{sec:discussion} investigates the possible physical contributions to the observed timing shift, including tidal and rotational deformation and perturbations from additional companions. Section \ref{sec:conclusion} summarises our findings and discusses prospects for future observations. The paper concludes with several appendices providing
details of the TESS, Spitzer, and JWST photometric analyses (Appendices~\ref{app:tess},
\ref{app:spitzer}, and
\ref{app:jwst}), as well as the results of our global
Newtonian orbital fit (Appendix~\ref{app:exofast}).
Appendix~\ref{app:gr} presents the theoretical derivation of the transit and eclipse timing corrections, including general-relativistic effects, adiabatic tidal deformations, and rotational effects.

\section{Data}
\label{sec: data}

\begin{table*}[t] 
\caption{Summary of the photometric observations of HD 80606 b used in this work.}
\label{tab:photometric_observations}
\centering
\begin{tabular}{lcc}
\hline\hline
Date & Telescope/instrument & Observation \\
\hline 
2007 Nov 20 & Spitzer/IRAC 8$\mu$m & Secondary eclipse \\
2009 Feb 12-14 & OHP 1.2 m & Partial transit \\
2010 Jan 13-14 & Spitzer/IRAC 4.5$\mu$m & Transit \\
2020 Feb 7 & TESS & Transit \\
2022 Nov 1-2 & JWST/NIRSpec NRS1 & Secondary eclipse \\
2022 Nov 1-2 & JWST/NIRSpec NRS2 & Secondary eclipse \\
2025 Apr 10-11 & JWST/MIRI LRS & Secondary eclipse \\
\hline
\end{tabular}
\end{table*}

We combined archival and publicly available photometric and radial-velocity observations of HD 80606 spanning more than two decades. The photometric data includes primary-transit observations from Spitzer and TESS, together with a partial primary transit observed by the Observatoire de Haute-Provence (OHP), and secondary-eclipse observations from Spitzer and JWST. Additionally, we use the archival radial-velocity measurements compiled by \citet{pearson2022utilizing}, supplemented by six HARPS-N measurements from \citet{figueira2016activity}. A summary of the datasets used in the global analysis is provided in Tables \ref{tab:photometric_observations} and \ref{tab:rv_observations}, with the individual datasets described below. 

\subsection{Ground-based Photometry}

The included ground-based photometry was presented by \citet{Moutou2009HD80606b}, obtained with the 1.2-m telescope at the OHP on 12-14 February 2009. The observations were acquired through a Bessel R filter and comprise a comparison night followed by partial coverage of the primary transit, including the full egress. We adopted the published differential photometry and converted the observation times from JD$_{UTC}$ to BJD$_{TDB}$. To account for the observed point-to-point scatter, the published photometric uncertainties were supplemented with a night-dependent noise floor added in quadrature, such that the median uncertainties matched the empirical scatter of each observing night. No additional detrending was applied before inclusion in the global fit. 

\subsection{TESS Photometry}
\label{sec:tess}
\begin{figure}
    \centering
    \includegraphics[width=1\linewidth]{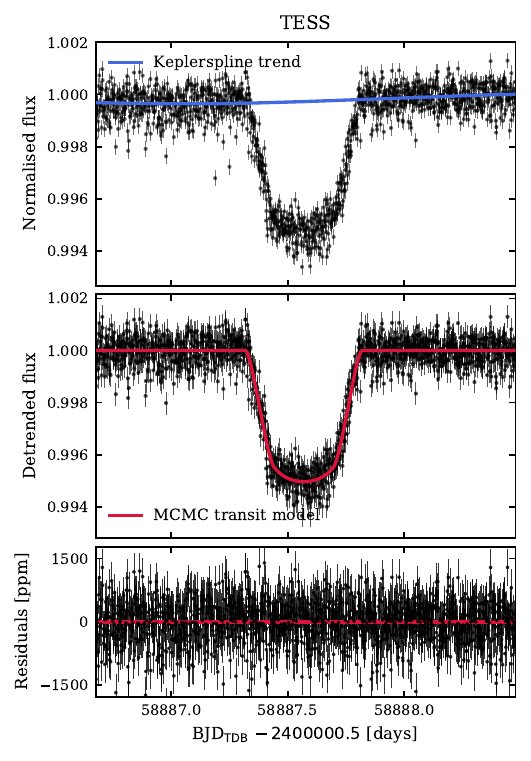}
    \caption[TESS transit light curve of HD 80606 b]{TESS primary-transit light curve of HD 80606 b. The upper panel shows the original normalised photometry together with the adopted \texttt{keplerspline} trend used to correct long-term variability. The middle panel shows the detrended light curve with the posterior-median transit model overplotted, and the lower panel shows the corresponding residuals.}
    \label{fig:tess_photometry}
\end{figure}

{ HD 80606 was observed by the Transiting Exoplanet Survey Satellite (TESS; \citealt{Ricker:2015}) at a cadence of 120~s during Sectors 21 and 47. Sector 21 covered a full primary transit of HD 80606 b, whereas the transit fell outside the observing window in Sector 47. We retrieved the TESS observations from the Mikulski Archive for Space Telescopes (MAST) using the \textit{Lightkurve} package \citep{Lightkurve:2018}. Owing to contamination from the nearby companion HD 80607, we extracted photometry using a custom aperture encapsulating both stars and subsequently detrended the light curve. The extraction, treatment of contamination, and detrending tests are described in Appendix~\ref{app:tess}. The resulting detrended transit light curve was included in the global fit described in Section~\ref{sec:updated_system_params}, with a dilution parameter accounting for the companion's flux. It is also presented in Fig.~\ref{fig:tess_photometry}}.

\subsection{Spitzer Photometry}
\begin{figure*}
\centering
\includegraphics[width=\textwidth]{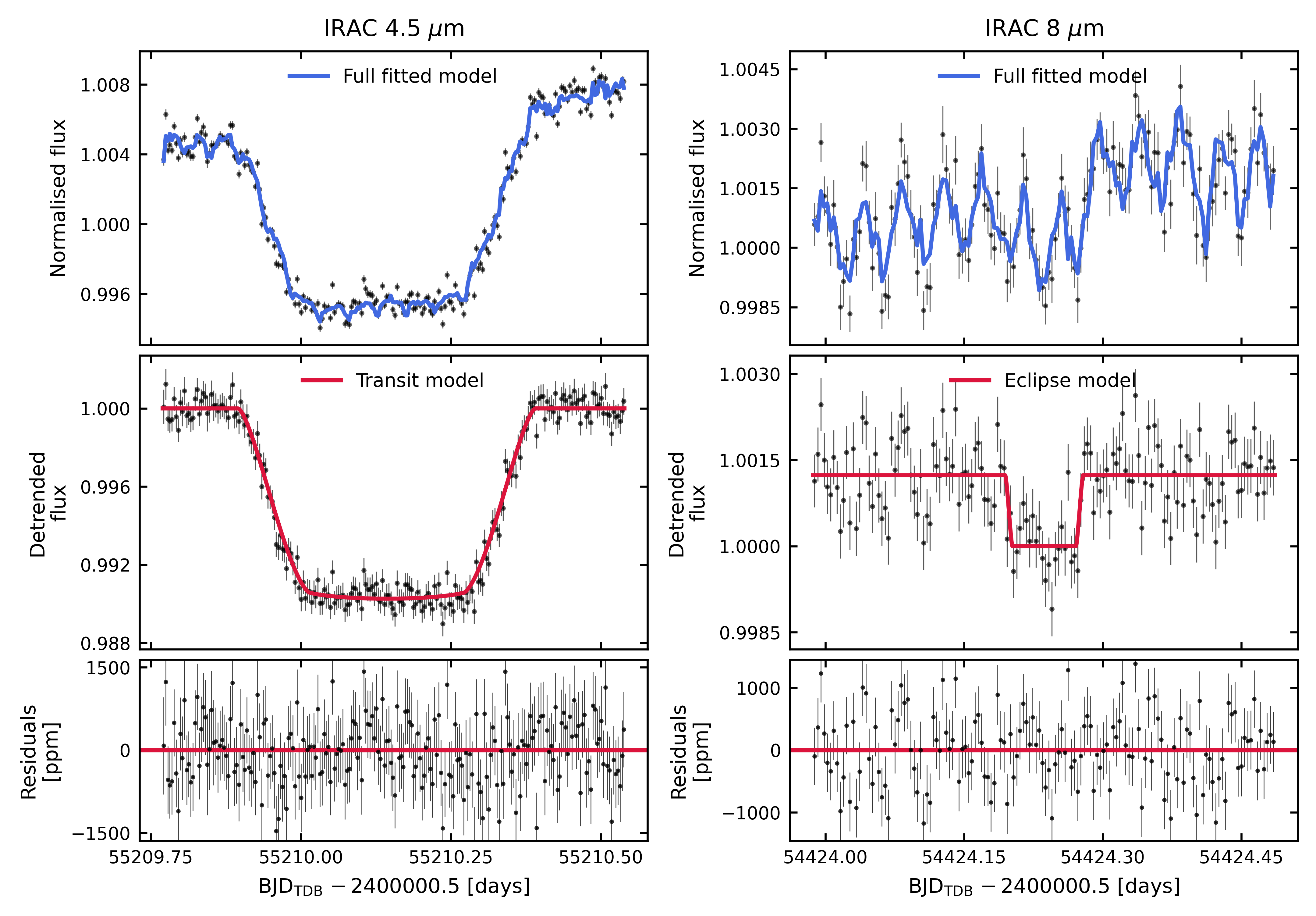}
\caption{Spitzer/IRAC observations of HD~80606~b at 4.5~$\mu$m
(left) and 8~$\mu$m (right). The upper panels show the normalised
photometry together with the full fitted models prior to detrending.
The middle panels show the detrended light curves with the transit
and secondary-eclipse models overplotted, respectively. The lower
panels show the corresponding residuals. The data are binned for
visualisation only; all fits were performed using the unbinned
photometry.}
\label{fig:spitzer}
\end{figure*}

{
We re-reduced archival Spitzer/IRAC observations of a primary transit and a secondary eclipse of HD 80606 b. The primary transit was observed at 4.5 $\mu$m on 13--14 January 2010 as part of Director's Discretionary Time program 540 and was originally presented by \citet{Hebrard2010HD80606b}. The secondary eclipse was observed at 8 $\mu$m on 20 November 2007 as part of program 40386 and was originally presented by \citet{Laughlin2009HD80606b}. The data reduction, modelling of pixel-phase and ramp effects, correction of instrumental systematics and the time-variable thermal emission of the planet, and fitting of the transit and eclipse light curves are described in the Appendix \ref{app:spitzer}. The uncorrected and corrected light curves, together with their best-fitting models, are presented in Fig.~\ref{fig:spitzer}.}

\subsection{JWST Photometry}\label{sec:jwst}
We retrieved publicly available, pipeline-processed JWST white-light curves of two secondary eclipses of HD 80606 b from the Mikulski Archive for Space Telescopes (MAST). The first observation was obtained on 1--2 November 2022 with NIRSpec as part of program GO-2488 (PI: J. Sikora; \citealt{Sikora2025HD80606b}), using the G395H grating and F290LP filter in bright-object time-series mode. Separate white-light curves were obtained for the NRS1 and NRS2 detectors. The second observation was obtained on 10--11 April 2025 with MIRI as part of program GO-2008 (PI: T. Kataria; \citealt{beichman2026eclipse}), using the low-resolution spectroscopy (LRS) slitless mode with the P750L prism.

\begin{figure*}
\centering
\includegraphics[width=\textwidth]{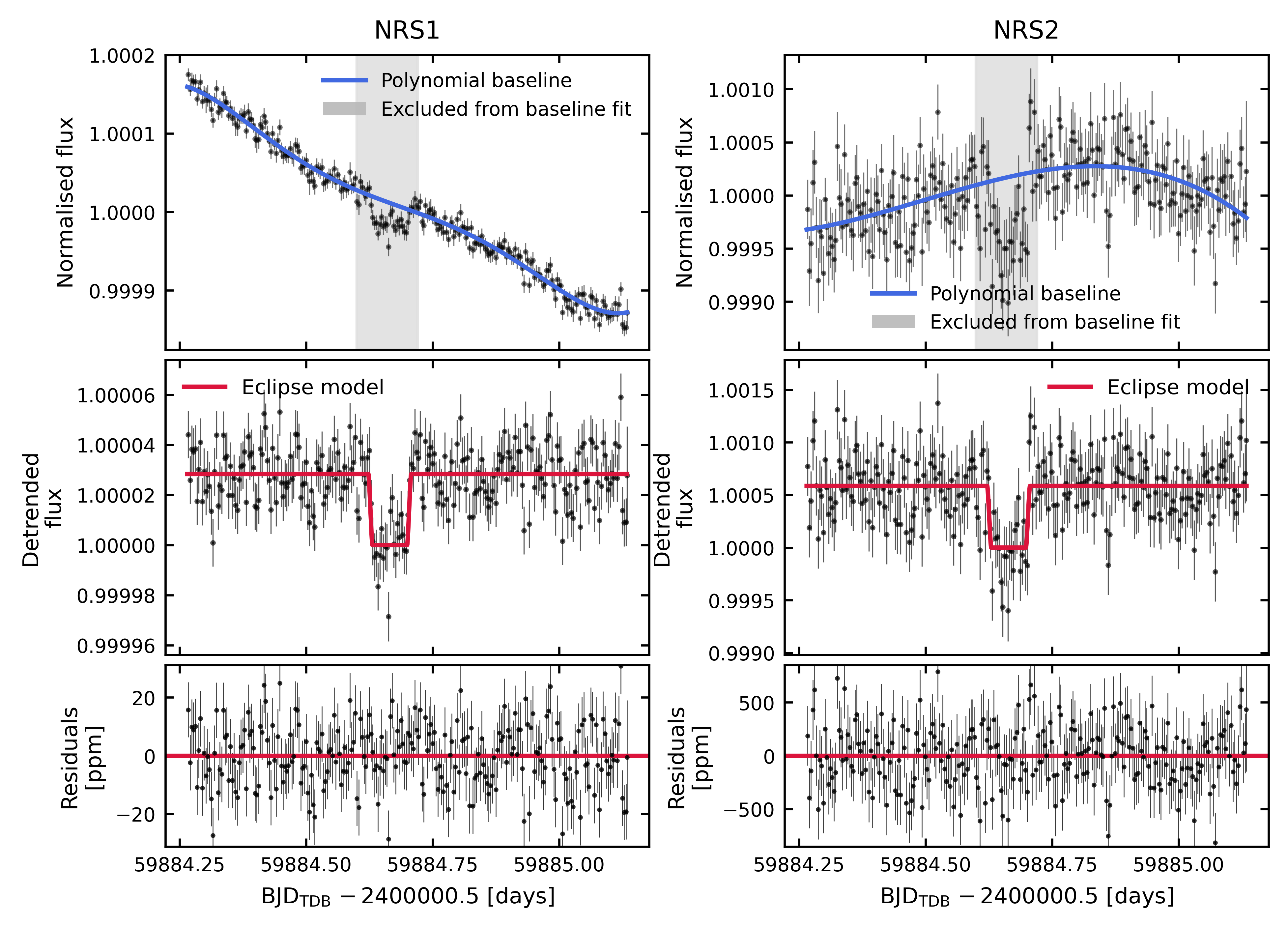}
\caption{Detrending and secondary-eclipse fits for the JWST/NIRSpec
observations of HD~80606~b. The left and right columns show the NRS1
and NRS2 detectors, respectively. The upper panels show the normalised
white-light curves together with the adopted polynomial baselines;
grey shading indicates regions excluded from the baseline fits. The
middle panels show the detrended light curves with the eclipse models
overplotted, and the lower panels show the corresponding residuals.
The data are binned for visualisation only; all fits were performed
using the unbinned photometry.}
\label{fig:jwst_nirspec}
\end{figure*}

\begin{figure}
\centering
\includegraphics[width=\columnwidth]{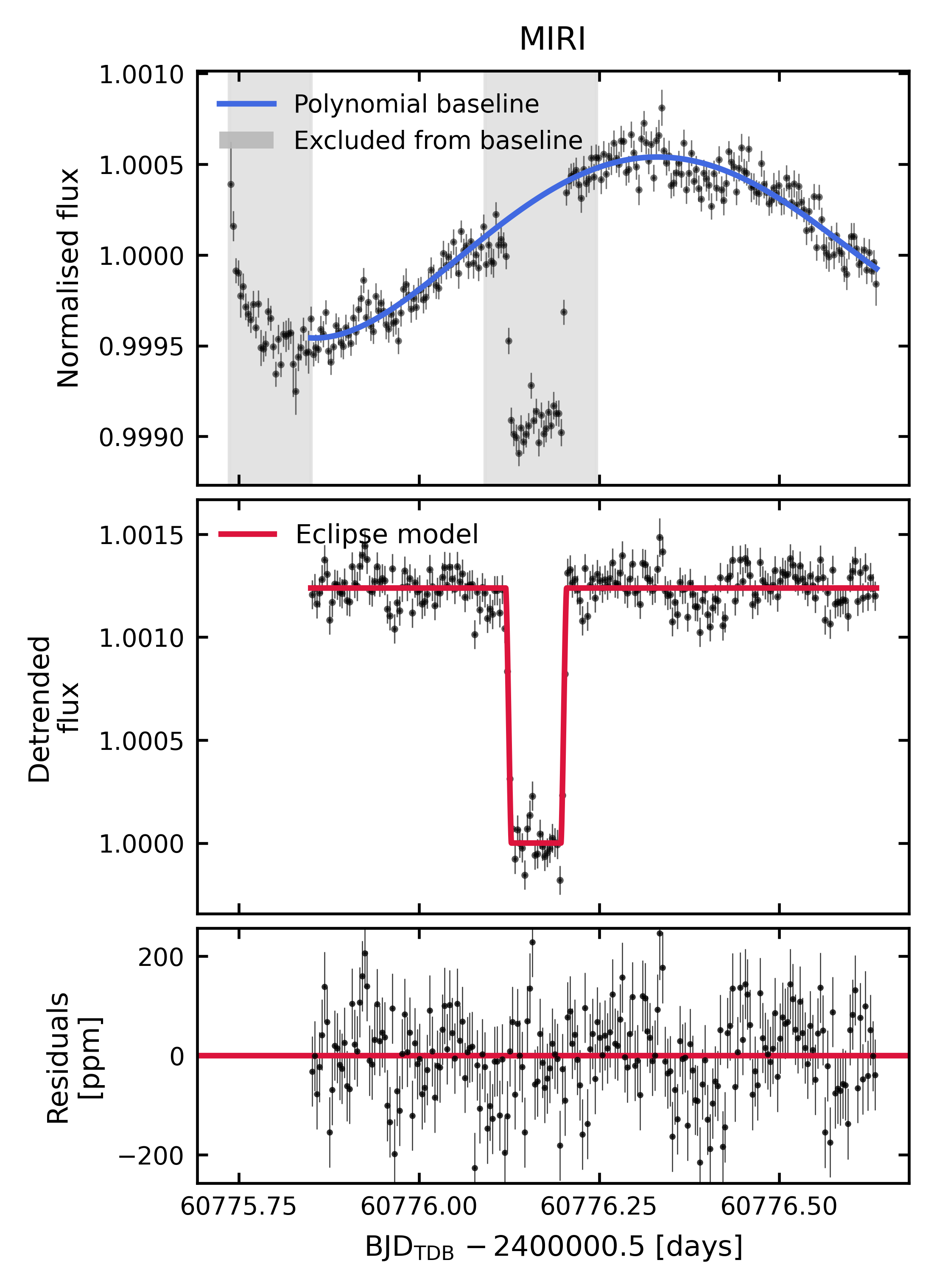}
\caption{Detrending and secondary-eclipse fit for the JWST/MIRI
observations of HD~80606~b. The upper panel shows the normalised
white-light curve together with the adopted polynomial baseline;
grey shading indicates regions excluded from the baseline fit.
The middle panel shows the detrended light curve with the eclipse
model overplotted, and the lower panel shows the corresponding
residuals. The data are binned for visualisation only; all fits were
performed using the unbinned photometry.}
\label{fig:jwst_miri}
\end{figure}

The archived white-light curves were produced from calibrated time-series spectroscopic observations using the JWST science calibration pipeline \citep{Bushouse2023JWST}. 
We applied additional corrections to account for residual instrumental systematics and the planet's phase-dependent thermal emission, as detailed in Appendix C. The original and corrected light curves, together with their best-fitting models, are presented in Figs.~\ref{fig:jwst_nirspec} and \ref{fig:jwst_miri}. The resulting unbinned, detrended light curves were then combined with the radial-velocity data in a global fit to refine the physical and orbital parameters of the system, as described in Sec.~\ref{sec:updated_system_params}.

\subsection{Radial-velocity observations}

We adopted the radial-velocity (RV) measurements compiled by \citet{pearson2022utilizing}, incorporating observations from ELODIE, Keck/HIRES, HET/HRS, SOPHIE, and APF. These measurements originate from the observing programmes described by \citet{Naef2001HD80606b}, \citet{butler2006catalog}, \citet{wittenmyer2007dynamical}, \citet{wittenmyer2009search}, \citet{Moutou2009HD80606b}, \citet{pont2009spin}, and \citet{rosenthal2021california}, together with additional HIRES and APF observations presented by \citet{pearson2022utilizing}, see Table \ref{tab:rv_observations}. We additionally included six HARPS-N radial-velocity observations published by \citet{figueira2016activity}. The combined measurements span approximately 22 years and were separated by instrument for use in the global \exofast analysis, with no additional processing applied.

\begin{table}
\caption{Summary of the radial-velocity observations
of HD 80606 used in this work.}
\label{tab:rv_observations}
\centering
\begin{tabular}{lccc}
\hline\hline
Instrument & Date range & $N_{\rm RV}$ & Ref. \\
\hline
ELODIE  & 1999--2003 & 74 & 1, 2 \\
HIRES K & 2001--2004 & 39 & 3 \\
HRS     & 2004--2007 & 23 & 4 \\
HIRES J & 2005--2021 & 75 & 5, 6 \\
SOPHIE  & 2009       & 48 & 7 \\
HARPS-N & 2016       & 6  & 8 \\
APF     & 2021       & 5  & 6 \\
\hline
\end{tabular}
\tablebib{
(1) \citet{Naef2001HD80606b};
(2) \citet{Moutou2009HD80606b};
(3) \citet{Winn2009HD80606b};
(4) \citet{wittenmyer2009search};
(5) \citet{rosenthal2021california};
(6) \citet{pearson2022utilizing};
(7) \citet{pont2009spin};
(8) \citet{figueira2016activity}.
}
\end{table}

\section{Updated system parameters}
\label{sec:updated_system_params}

Prior to performing a detailed timing analysis, we updated the global parameters for the HD 80606 system. Running an independent fit ensures self-consistency in the parameters that we subsequently fix when analysing the individual transit and eclipse events. We used the exoplanet fitting software \exofast \citep{Eastman:2013,Eastman:2017,Eastman:2019} to obtain these updated system parameters. \exofast utilises a differential evolution Markov Chain Monte Carlo (DE-MCMC) algorithm to fit stellar and planetary parameters simultaneously. 

 We provide \exofast\ with the photometric light curves and
radial-velocity datasets used in this study, together with
the broadband spectral energy distribution (SED) of HD 80606,
and impose priors on a limited set of model parameters.
The datasets and adopted priors are described in detail below.

We placed priors on several stellar parameters. The line-of-sight V-band extinction $A_V$ was constrained with a uniform prior of $U[0.0,0.043]$ mag, with the upper bound being a conservative limit from \cite{Schlafly:2011}. We implemented a Gaussian prior on the stellar metallicity  which was derived by \citet{rosenthal2021california}. Finally, a Gaussian prior was also placed on the parallax, which was taken from Gaia Data Release 3 (DR3; \citealt{gaiadr3}). According to \cite{Lindegren:2021}, a small correction ($\sim$0.03 mas) was applied to the parallax to correct for the zero-point offset, and the uncertainty was added in quadrature with 0.01 to capture remaining systematics. Thus, the final parallax prior was $15.17413 \pm 0.01972$ mas. 

The stellar SED was constructed using broadband photometry from Gaia DR3 \citep{gaiadr3}, the Two Micron All Sky Survey (2MASS; \citealt{Cutri:2003}) and the Wide-field Infrared Survey Explorer (WISE; \citealt{Cutri:2012}). To constrain the host star, we use the default setting within \exofast using the MESA Isochrones and Stellar Tracks (MIST) stellar evolution models \citep{Paxton:2011,Paxton:2013,Paxton:2015,Dotter:2016,Choi:2016}. \exofast fits the quadratic limb darkening by interpolating the models from \cite{Claret:2011}. 

As mentioned in Section \ref{sec:tess}, the nearby companion HD 80607 introduced significant contamination to the TESS aperture. As we selected a custom aperture that contained the flux from both stars in the system, we enforced a dilution term for the TESS transit within the \exofast fit in order to offset the contamination. We placed a conservative uniform prior of $U[0,0.56]$ on the dilution, which in \exofast is defined as $F_2/(F_1 + F_2)$, where $F_1$ and $F_2$ are the flux from the primary and secondary star, respectively.

The \exofast fit was accepted as converged once the Gelman-Rubin metric fell below 1.03 and the number of independent draws, $T_z$, exceeded 1000. The resulting planetary and orbital parameters are presented in Table~\ref{tab:planet_parameters}, together with the corresponding published values from \citet{pearson2022utilizing}. The fitted stellar parameters are similarly compared with published measurements in Table~\ref{tab:stellar_parameters}. The parameters obtained from the global fit were fixed during the subsequent timing analysis, in which only the event midpoint and transit or eclipse depth were allowed to vary. The full \exofast tables are shown in Appendix \ref{app:exofast}.

\begin{table*}
\caption{
Comparison of published parameters for HD 80606 b
from \citet{pearson2022utilizing} with posterior median values
derived from our \exofast global fit.
The quoted \exofast uncertainties represent
the 68\% credible intervals.
}
\label{tab:planet_parameters}
\centering
\begin{tabular}{lcc}
\hline\hline
Parameter & \citet{pearson2022utilizing} & This work \\
\hline
$P$ (d)
& $111.436765 \pm 0.000074$
& $111.437252 \pm 0.000022$ \\

$a$ (AU)
& $0.4603 \pm 0.0021$
& $0.4551^{+0.0069}_{-0.0068}$ \\

$e$
& $0.93183 \pm 0.00014$
& $0.93116 \pm 0.00030$ \\

$i$ ($^\circ$)
& $89.24 \pm 0.01$
& $89.2289^{+0.0071}_{-0.0070}$ \\

$b$ ($R_\star$) 
& ... 
& $0.8155^{+0.0036}_{-0.0037}$ \\

$\omega_\star$ ($^\circ$)
& $-58.887 \pm 0.043$
& $-58.668 \pm 0.085$ \\

$M_{\rm p}$ ($M_{\rm Jup}$)
& $4.1641 \pm 0.0047$
& $4.09^{+0.13}_{-0.12}$ \\

$R_{\rm p}$ ($R_{\rm Jup}$)
& $1.032 \pm 0.015$
& $1.011^{+0.017}_{-0.016}$ \\

$T_{\rm C,obs}$ (BJD$_{\rm TDB}$)
& $2458888.07466 \pm 0.00204$
& $2458888.07964^{+0.00079}_{-0.00080}$ \\

$T_{\rm P}$ (BJD$_{\rm TDB}$)
& $2458882.3440 \pm 0.0021$
& $2458882.3618^{+0.0014}_{-0.0013}$ \\

$\dot{\omega}_{\rm GR, pred}$ ($^\circ$ century$^{-1}$)
& \ldots
& $0.0583^{+0.0018}_{-0.0017}$ \\
\hline
\end{tabular}
\end{table*}

\begin{table*}
\caption{Comparison between adopted published stellar parameters for HD~80606 and posterior median values obtained from the \exofast global fit. Published sources are indicated individually. \exofast uncertainties correspond to the 68\% credible interval.}
\label{tab:stellar_parameters}
\centering
\begin{tabular}{lccc}
\hline\hline
Parameter & published value & Source & This work \\
\hline
$T_{\rm eff}$ (K)
& $5566 \pm 92$
& \citet{rosenthal2021california}
& $5541^{+51}_{-50}$ \\

$\log g$ (cgs)
& $4.402 \pm 0.032$
& \citet{rosenthal2021california}
& $4.4001^{+0.0086}_{-0.0086}$ \\

$[\mathrm{Fe/H}]$ (dex)
& $0.348 \pm 0.058$
& \citet{rosenthal2021california}
& $0.352^{+0.074}_{-0.076}$ \\

$M_\star$ ($M_\odot$)
& $1.018 \pm 0.035$
& \citet{bonomo2017gaps}
& $1.009^{+0.047}_{-0.045}$ \\

$R_\star$ ($R_\odot$)
& $1.037 \pm 0.032$
& \citet{bonomo2017gaps}
& $1.049^{+0.016}_{-0.016}$ \\

Distance (pc)
& $66.03 \pm 0.07$
& \citet{gaiadr3}
& $65.900^{+0.086}_{-0.085}$ \\
\hline
\end{tabular}
\end{table*}

\begin{figure*}[!htbp]
    \centering
    \includegraphics[width=0.49\textwidth]{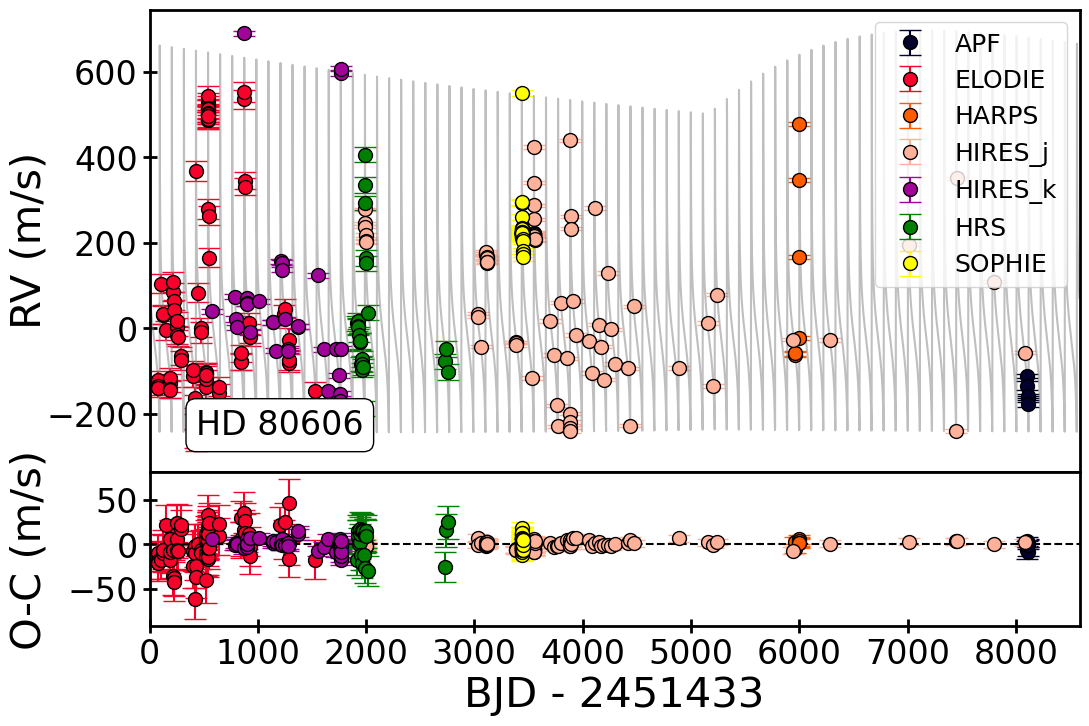}
    \hfill
    \includegraphics[width=0.49\textwidth]{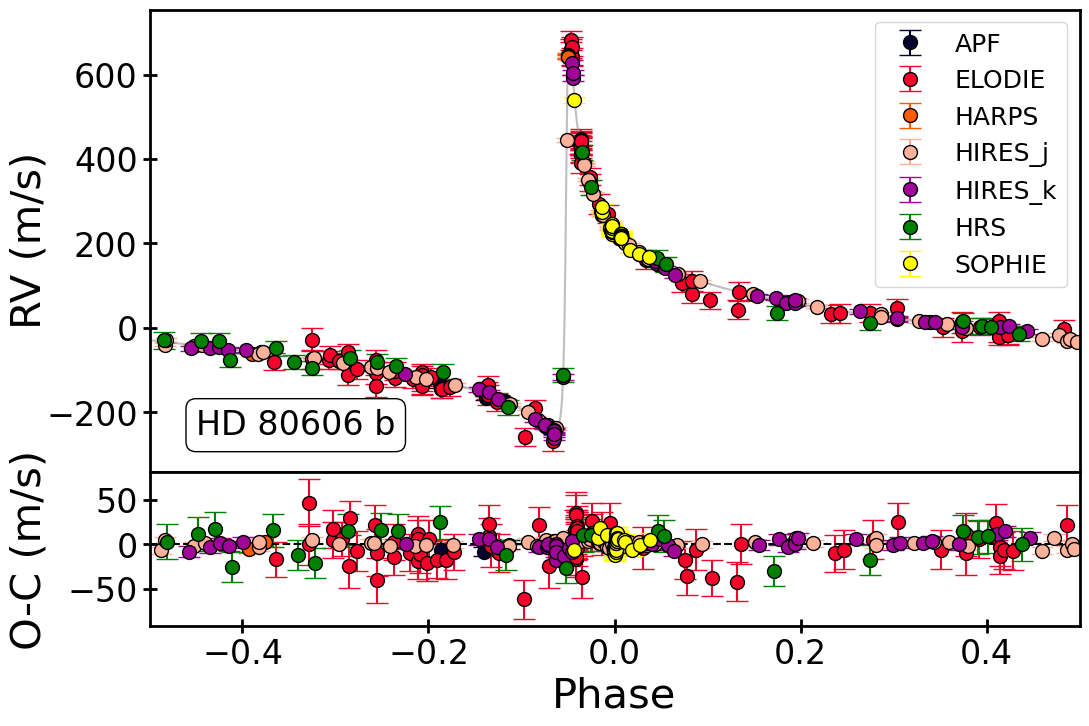}

    \caption[Radial-velocity observations and orbital fit for HD 80606 b]{Radial-velocity observations of HD 80606 from seven instrumental datasets, together with the best-fitting orbital model. The left panel shows the observations as a function of time, while the right panel presents the phase-folded radial-velocity observations. The lower panels show the observed-minus-calculated (O--C) residuals.}
    \label{fig:rv_fit}
\end{figure*}

The radial-velocity observations and the corresponding best-fitting \exofast orbital model are shown in Fig. \ref{fig:rv_fit}, both as a function of time and folded by orbital phase, together with the observed-minus-calculated residuals.

\section{Predicting the transit--eclipse separation}
\label{sec:theoretical predictions}

To interpret the evolution of the transit and secondary-eclipse timings of HD~80606~b, we revisited the theoretical framework developed by \citet{blanchet2019detecting}, using the updated physical and orbital parameters obtained from our global fit (Section~\ref{sec:updated_system_params}). This framework describes the orbital motion through a quasi-Keplerian parametrisation incorporating general-relativistic corrections and Newtonian tidal and rotational deformations of the star and planet. The detailed formulation, including the perturbations to the orbital elements and the calculation of the corresponding event times, is presented in Appendix~\ref{app:gr}. The result is the relativistic time shifts for transits denoted
\begin{align}\label{deltatindiv}
	&\bigl\{\delta t_1(N), \delta t_2(N),\delta t_m(N), \delta t_3(N),\delta t_4(N)\bigr\},
\end{align}
where $N$ is the number of the orbit, and the labels $i \in \{1,2,m,3,4\}$ correspond respectively to the entry of the transit, the start of the full transit, the midpoint, the exit of the full transit and exit of the partial transit. Similarly for the time shifts of eclipses denoted with an overbar,
\begin{align}\label{deltatbarindiv}
	&\bigl\{\delta \bar{t}_1(N), \delta \bar{t}_2(N),\delta \bar{t}_m(N), \delta \bar{t}_3(N),\delta \bar{t}_4(N)\bigr\}.
\end{align}
The time shifts \eqref{deltatindiv}--\eqref{deltatbarindiv} are defined as the differences
\begin{equation}\label{titibar}
    \delta t_i(N) = t_i(N) - t^0_i(N),\qquad \delta \bar{t}_i(N) = \bar{t}_i(N) - \bar{t}{}^0_i(N),
\end{equation}
between the real physical model and the reference Newtonian model obtained from our global fit in Sec.~\ref{sec:updated_system_params}. The time shifts are reported in Table~\ref{tab:gr_individual_shifts} for the pure GR effect and a number of past and future transits and eclipses.

The individual shifts \eqref{titibar} are not directly observable because they depend on the reference Newtonian model. Following \citet{blanchet2019detecting}, we instead consider a combination of transit and eclipse times that removes this dependence. Let us define the time interval between the transit of the $N$-th orbit and the eclipse of the $M$-th orbit, \textit{i.e.}
\begin{equation}\label{ttrec}
t_\text{tr-ec}(N,M) = t_m(N) - \bar{t}_m(M),
\end{equation}
where $t_m(N)$ is the mid-transit time and $\bar{t}_m(M)$ is the mid-eclipse time.\footnote{In our convention the orbit $N$ contains the $N$-th eclipse at instant $\bar{t}_m(N)$ followed by the $N$-th transit at instant $t_m(N)$. The orbit $N$ starts from the apastron which occurred before the eclipse $N$, and ends at the next apastron after the transit $N$ (remembering that the periastron of the orbit is in between the eclipse and the following transit). Following \citet{blanchet2019detecting} the orbit $N=0$ corresponds to the January 2010 transit.} The time interval \eqref{ttrec} can be positive or negative and we define
\begin{equation}\label{tectr}
t_\text{ec-tr}(M,N) = - t_\text{tr-ec}(N,M).
\end{equation}
As the transit and eclipse observations may correspond to different orbital cycles, we subtract the elapsed integer number of orbital periods to obtain a separation that can be compared across epochs.
We therefore introduce the ``period-corrected'' separation
\begin{equation}\label{ttreccorr}
t^\text{corr}_\text{tr-ec}(N,M) \equiv  t_\text{tr-ec}(N,M) - (N-M)P,
\end{equation}
where $P$ is the orbital period from the global fit in Sec.~\ref{sec:updated_system_params}.
To investigate the evolution of the transit--eclipse separation, we compare~\eqref{ttreccorr} with another measurement of a transit--eclipse separation corresponding to different orbits $N'$ and $M'$, preferably well separated from the orbits $N$ and $M$. We therefore define the observed timing combination as
\begin{align}\label{Deltatobs}
\Delta t^\text{obs} \equiv t^\text{corr}_\text{tr-ec}(N',M') - t^\text{corr}_\text{tr-ec}(N,M).
\end{align}
A straightforward generalization of Eq.~(30) in \citet{blanchet2019detecting}, using the fact that the Newtonian orbit is fixed, thus $t^0_i(N) = t^0_i(0) + N P$ and $\bar{t}{}^0_i(N) = \bar{t}{}^0_i(0) + N P$, gives 
\begin{align}\label{predictionGR}
	\Delta t^\text{obs} = \Delta t^\text{pred},
\end{align}
where the prediction $\Delta t^{\text{pred}}$ is the following combination of individual shifts 
\begin{align}\label{GRprediction}
	\Delta t^\text{pred} = \delta \bar{t}_m(M) - \delta t_m(N) - \delta \bar{t}_m(M') + \delta t_m(N') .
\end{align}
In the absence of apsidal precession and bodies' internal structure, this relativistic timing combination vanishes.

General relativity introduces a non-zero contribution, which can be computed (at the relativistic level 1PN) from the perturbations to the individual event times which are listed in Table~\ref{tab:gr_individual_shifts}. The same perturbative framework also allows us to calculate the contributions from Newtonian tidal and rotational deformations, which depend on the Love numbers of the star and planet. Hence we consider
\begin{align}\label{GR+}
    \Delta t^\text{pred} = \Delta t^{\text{GR}} + \Delta t^{\text{tide}} + \Delta t^{\text{rot}}.
\end{align}
The underlying calculations are described in Appendix~\ref{app:gr}, while the physical interpretation of these additional contributions is discussed in Section~~\ref{sec:discussion}.

The theoretical predictions for the timing combinations $\Delta t^\text{pred}$ of past and future observations (including those scheduled for Ariel) are given in Table~\ref{tab:gr_only_predictions} for GR alone, and in Table~\ref{tab:gr_tides_kstar} when tidal and rotational contributions are included.
\begin{table*}[t]
	\begin{center}
		\begin{tabular}{| l || c | c | c |c | c |}
			\hline\hline
			number $N$ of transit & $\delta t_1(N)$ & $\delta t_2(N)$ & $\delta
			t_m(N)$ & $\delta t_3(N)$ & $\delta t_4(N)$ \\
			\hline
			0 (January 2010) & $-2.61$ & $-2.65$ & $-2.69$ & $-2.72$ & $-2.76$ \\
			1 (May 2010) & $-7.58$ & $-7.70$ & $-7.81$ & $-7.93$ & $-8.05$ \\
			\hspace{1cm} $\vdots$ & $\vdots$ & $\vdots$ & $\vdots$ & $\vdots$ & $\vdots$ \\
			33 (February 2020) &  $-166.6$ & $-169.2$ & $-171.8$ & $-174.4$ & $-177.1$ \\
			\hspace{1cm} $\vdots$ & $\vdots$ & $\vdots$ & $\vdots$ & $\vdots$ & $\vdots$ \\
			52 (25 November 2025) &  $-261.0$ & $-265.2$ & $-269.2$ & $-273.3$ & $-277.4$ \\
			53 (17 March 2026) &  $-266.0$ & $-270.2$ & $-274.4$ & $-278.5$ & $-282.7$ \\
			54 (6 July 2026) &  $-271.0$ & $-275.3$ & $-279.5$ & $-283.7$ & $-288.0$ \\
			55 (26 October 2026) &  $-275.9$ & $-280.3$ & $-284.6$ & $-288.9$ & $-293.3$ \\
			\hspace{1cm} $\vdots$ & $\vdots$ & $\vdots$ & $\vdots$ & $\vdots$ & $\vdots$ \\
			76 (23 March 2033) &  $-380.3$ & $-386.3$ & $-392.3$ & $-398.2$ & $-404.2$ \\
			\hspace{1cm} $\vdots$ & $\vdots$ & $\vdots$ & $\vdots$ & $\vdots$ & $\vdots$ \\
			86 (10 April 2036) &  $-430.0$ & $-436.8$ & $-443.5$ & $-450.2$ & $-457.0$ \\
			\hspace{1cm} $\vdots$ & $\vdots$ & $\vdots$ & $\vdots$ & $\vdots$ & $\vdots$ \\
			99 (29 March 2040) &  $-494.6$ & $-502.4$ & $-510.2$ & $-517.9$ & $-525.7$ \\
			\hline
		\end{tabular}
	\end{center}\begin{center}
		\begin{tabular}{| l || c | c | c |c | c|}
			\hline\hline
			number $N$ of eclipse & $\delta \bar{t}_1(N)$ & $\delta \bar{t}_2(N)$ & $\delta
			\bar{t}_m(N)$ & $\delta \bar{t}_3(N)$ & $\delta \bar{t}_4(N)$ \\
			\hline
			$-$7 (November 2007) & $-2.24$ & $-2.25$ & $-2.26$ &
			$-2.28$ & $-2.28$ \\
			\hspace{1cm} $\vdots$ & $\vdots$ & $\vdots$ & $\vdots$ & $\vdots$ & $\vdots$ \\
			0 (January 2010) & $8.5\cdot10^{-3}$ & $8.5\cdot10^{-3}$ & $8.9\cdot10^{-3}$ &
			$9.4\cdot10^{-3}$ & $9.5\cdot10^{-3}$ \\
			1 (April 2010) & $0.33$ & $0.33$ & $0.33$ & $0.34$ & $0.34$ \\
			\hspace{1cm} $\vdots$ & $\vdots$ & $\vdots$ & $\vdots$ & $\vdots$ & $\vdots$ \\
			42 (1 November 2022) &  $13.52$ & $13.54$ & $13.65$ & $13.75$ & $13.77$ \\
			\hspace{1cm} $\vdots$ & $\vdots$ & $\vdots$ & $\vdots$ & $\vdots$ & $\vdots$ \\
			50 (11 April 2025) &  $16.09$ & $16.12$ & $16.24$ & $16.37$ & $16.40$ \\
			\hspace{1cm} $\vdots$ & $\vdots$ & $\vdots$ & $\vdots$ & $\vdots$ & $\vdots$ \\
			76 (17 March 2033) &  $24.45$ & $24.50$ & $24.69$ & $24.88$ & $24.92$ \\
			\hspace{1cm} $\vdots$ & $\vdots$ & $\vdots$ & $\vdots$ & $\vdots$ & $\vdots$ \\
			86 (4 April 2036) &  $27.67$ & $27.72$ & $27.93$ & $28.15$ & $28.20$ \\
			\hspace{1cm} $\vdots$ & $\vdots$ & $\vdots$ & $\vdots$ & $\vdots$ & $\vdots$ \\
			99 (23 March 2040) &  $31.85$ & $31.91$ & $32.15$ & $32.40$ & $32.46$ \\
			\hline
		\end{tabular}
		\caption{Predicted relativistic shifts (in seconds) of transit (upper table) and eclipse (lower table) contact times computed with the relativistic model (including the GR effect but with no tides and rotational deformations), taking as reference point (where $\delta t=\delta \bar{t} =0$) the passage at periastron during the reference orbit $N=0$.}\label{tab:gr_individual_shifts}
	\end{center}
\end{table*}

\begin{table*}[t]
	\begin{center}
		\begin{tabular}{| c || c |}
			\hline\hline
			$(M, N)$ 
			& $\begin{array}{c} \quad\text{predicted GR effect}\\[0.0cm] \Delta t^\text{GR} \text{(without tides and rotation)}
			\end{array}$ \\
			\hline
			$(42, 33)$ & $185.07$ \\
			\hline
			$(50, 33)$ & $187.67$ \\
			\hline
			$(76,76)$ & $416.53$ \\
			\hline
			$(86,86)$ & $471.04$ \\
			\hline
			$(99,99)$ & $541.90$ \\
			\hline
		\end{tabular}
	\end{center}
	\caption{Predicted GR-only timing shifts $\Delta t^\text{GR}$ (in seconds) for the TESS transit ($N=33$) and the JWST eclipses ($M=42,50$). We include also future Ariel observations with $N=M=$ 76, 86, 99. We use the Spitzer reference transit--eclipse interval~\eqref{refinterval}.}\label{tab:gr_only_predictions}
\end{table*}
\begin{table*}[t]
	\begin{center}
		\begin{tabular}{| l || c | c | c |c | c |}
\hline\hline
& \multicolumn{5}{c|}{$\Delta t^\text{pred} = \Delta t^\text{GR} + \Delta t^\text{tide} + \Delta t^\text{rot}$} \\ 
\cline{2-6} 
$(M,N)$
& $\begin{array}{c}
k_{\mathrm{p}}=0\\
k_\star=0
\end{array}$
& $\begin{array}{c}
k_{\mathrm{p}}=0.2\\
k_\star=0
\end{array}$
& $\begin{array}{c}
k_{\mathrm{p}}=0.4\\
k_\star=0
\end{array}$
& $\begin{array}{c}
k_{\mathrm{p}}=0.6\\
k_\star=0
\end{array}$
& $\begin{array}{c}
k_{\mathrm{p}}=0.8\\
k_\star=0
\end{array}$ \\
\hline
$(42,33)$ & $185.07$ & $207.47$ & $229.88$ & $252.28$ & $274.69$ \\
\hline
$(50,33)$ & $187.67$ & $210.05$ & $232.42$ & $254.80$ & $277.18$ \\
\hline
$(76,76)$ & $416.53$ & $468.22$ & $519.92$ & $571.61$ & $623.30$ \\
\hline
$(86,86)$ & $471.04$ & $529.54$ & $588.03$ & $646.53$ & $705.03$ \\
\hline
$(99,99)$ & $541.90$ & $609.24$ & $676.59$ & $743.93$ & $811.27$ \\
\hline	\end{tabular}
	\end{center}\begin{center}
		\begin{tabular}{| l || c | c | c |c | c |}
\hline\hline& \multicolumn{5}{c|}{$\Delta t^\text{pred} = \Delta t^\text{GR} + \Delta t^\text{tide} + \Delta t^\text{rot}$} \\ 
\cline{2-6} 
$(M,N)$
& $\begin{array}{c}
k_{\mathrm{p}}=0\\
k_\star=0.2
\end{array}$
& $\begin{array}{c}
k_{\mathrm{p}}=0.2\\
k_\star=0.2
\end{array}$
& $\begin{array}{c}
k_{\mathrm{p}}=0.4\\
k_\star=0.2
\end{array}$
& $\begin{array}{c}
k_{\mathrm{p}}=0.6\\
k_\star=0.2
\end{array}$
& $\begin{array}{c}
k_{\mathrm{p}}=0.8\\
k_\star=0.2
\end{array}$ \\
\hline
$(42,33)$ & $221.60$ & $244.01$ & $266.41$ & $288.81$ & $311.22$ \\
\hline
$(50,33)$ & $224.15$ & $246.53$ & $268.91$ & $291.29$ & $313.67$ \\
\hline
$(76,76)$ & $500.83$ & $552.52$ & $604.22$ & $655.91$ & $707.60$ \\
\hline
$(86,86)$ & $566.43$ & $624.93$ & $683.43$ & $741.93$ & $800.43$ \\
\hline
$(99,99)$ & $651.72$ & $719.07$ & $786.41$ & $853.75$ & $921.10$ \\
\hline
	\end{tabular}
	\caption{Predicted $\Delta t^\text{pred} = \Delta t^\text{GR} + \Delta t^\text{tide} + \Delta t^\text{rot}$ timing shifts (in seconds) for $k_\star=0$ (upper panel) and $k_\star=0.2$ (lower panel), and for different planetary Love numbers $k_{\mathrm{p}}$ (we pose $k_\star\equiv k_\star^{(2)}$and $k_p\equiv k_p^{(2)}$). The star's rotation period is $P_\star \simeq 40\,\mathrm{d}$ and the planet's rotation period is $P_{\mathrm{p }}\simeq 93\,\mathrm{h}$. We use the Spitzer reference transit--eclipse interval~\eqref{refinterval}.}\label{tab:gr_tides_kstar}
	\end{center}
\end{table*}

\section{Timing measurements}
\label{sec: timing measurements}

In this section, we first derive independent transit and secondary-eclipse midpoints from the available photometric observations, using the orbital parameters obtained from
our global fit.
We then combine these measurements following the formalism of Sec.~\ref{sec:theoretical predictions} to search for a secular evolution of the transit--eclipse timing separation and compare the observed signal with theoretical predictions.

\subsection{Individual timing measurements}

To obtain independent measurements of the transit and secondary-eclipse midpoints, each detrended light curve was fitted individually using the \texttt{emcee} Markov Chain Monte Carlo (MCMC) sampler \citep{emcee}. The orbital geometry was fixed to the posterior median values obtained from the global \exofast fit described in Section \ref{sec:updated_system_params}, including the orbital period, scaled semi-major axis, inclination, eccentricity, argument of periastron, and planet-to-star radius ratio. Each individual fit therefore contained only two free parameters: the event midpoint and observed transit or eclipse depth. For transits, the observed depth was varied by scaling the amplitude of a fixed-shape transit model, while for secondary eclipses, the depth was fitted through the planet-to-star flux ratio. In both cases, the orbital geometry, including the planet-to-star radius ratio, remained fixed. Transit and eclipse models were generated using \texttt{batman} \citep{batman:2015}, with fixed four-parameter limb-darkening coefficients obtained from ExoCTK \citep{matthew_bourque_2021_4556063} for the corresponding instrumental bandpasses. The Cousins R bandpass was adopted as a proxy for the OHP Bessel R observations. Limb darkening is not required for the secondary-eclipse models. 

Uniform priors were adopted for the midpoint and depth, with the midpoint restricted to within $\pm\,0.05$ d of its expected value. Each fit used 48 walkers evolved for 5000 steps, with the first 1000 steps discarded as burn-in and the chains thinned by a factor of 10. The median of the marginalised midpoint posterior was adopted as the measured event time, with uncertainties corresponding to the 16th and 84th percentiles. The resulting timings are presented in Table \ref{tab:timings}. The TESS photometric uncertainties were rescaled by a common factor to account for the underestimated scatter in the light curve, such that the final fit yielded a reduced $\chi^2$ close to unity.

\begin{table*}
\caption{Transit and secondary-eclipse timings measured in this work. Orbit numbers follow the convention of \citet{blanchet2019detecting}, with \(N=0\) corresponding to the January 2010 transit.}
\label{tab:timings}
\centering
\begin{tabular}{lcccc}
\hline\hline
Telescope/instrument & Event & Orbit $N$ &
\shortstack{$T_{\rm mid}$ \\ $(\mathrm{BJD}_{\rm TDB}-2400000.5)$} & $\sigma_{T_{mid}}$ [min]\\
\hline
Spitzer/IRAC 8~$\mu$m
& Eclipse
& $-7$
& $54424.23667503$
& $^{+1.67}_{-1.83}$ \\

OHP 1.2~m
& Transit
& $-3$
& $54875.83045860$
& $^{+5.37}_{-5.44}$ \\

Spitzer/IRAC 4.5~$\mu$m
& Transit
& $0$
& $55210.15189671$
& $^{+1.29}_{-1.27}$ \\

TESS
& Transit
& $33$
& $58887.57786962$
& $^{+1.24}_{-1.19}$ \\

JWST/NIRSpec NRS1
& Eclipse
& $42$
& $59884.66525640$
& $^{+1.46}_{-1.43}$ \\

JWST/NIRSpec NRS2
& Eclipse
& $42$
& $59884.66416716$
& $^{+1.56}_{-1.70}$ \\

JWST/MIRI
& Eclipse
& $50$
& $60776.16257419$
& $^{+0.30}_{-0.30}$ \\
\hline
\end{tabular}
\end{table*}

The individual fits provide timing measurements spanning approximately 17 years, from the 2007 Spitzer eclipse to the 2025 JWST/MIRI eclipse. The timing precision varies substantially between datasets, from approximately 5.4 min for the partial OHP transit to 0.30 min for the MIRI eclipse. The independently fitted NIRSpec NRS1 and NRS2 light curves yield eclipse midpoints differing by only $\sim$\,1.6 min and are consistent within their uncertainties. 

The measured event times are generally consistent with previous determinations from the corresponding observations \citep{Laughlin2009HD80606b, Moutou2009HD80606b, pearson2022utilizing, Sikora2025HD80606b, beichman2026eclipse}. In particular, the NIRSpec eclipse times differ from those reported by \citet{Sikora2025HD80606b} by only a few seconds, while the MIRI eclipse time is consistent with the independent processing of \citet{beichman2026eclipse}. The notable exception is the January 2010 Spitzer transit, for which we obtain \(T_{\rm mid}=2455210.65190\) BJD\(_{\rm TDB}\), approximately 14 min later than the value reported by \citet{Hebrard2010HD80606b}. To test whether this discrepancy resulted from our independent reduction of the Spitzer observations, we repeated our timing analysis using the reduced light curve published by \citet{Hebrard2010HD80606b}. This yielded $T_{\mathrm{mid}} = 2455210.65090$ BJD, consistent with the value obtained from our independently reduced photometry. Furthermore, the independent ground-based measurement of the same transit by \citet{shporer2010ground}, $T_{\mathrm{mid}} = 2455210.6502 \pm 0.0064$, is also consistent with our result. The discrepancy with the published \citet{Hebrard2010HD80606b} midpoint is therefore unlikely to arise from differences in the photometric reduction alone. It may instead reflect the orbital solution adopted by \citet{Hebrard2010HD80606b}, which was derived from a substantially shorter observational baseline than the global fit used here. 

\subsection{ Timing evolution and apsidal precession}

We searched for the predicted evolution of the transit--eclipse separation using the timing combination defined in Eq.~\eqref{GRprediction}. The reference interval was determined from the re-analysis of the Spitzer eclipse at orbit $M'=-7$ and the subsequent Spitzer transit at $N'=0$. The result of this re-analysis is the measurement of the reference period-corrected interval, following the notation~\eqref{ttreccorr}, as
\begin{equation}\label{refinterval}
t^\text{corr}_\text{tr-ec}(0,-7) = t_m(0) -\bar{t}_m(-7) - 7 P = 5.85445768\, \mathrm{d}.
\end{equation}
For a later transit at orbit $N$ and eclipse at orbit $M$, the observed relativistic timing combination is 
\begin{align}\label{obsmoins7}
\Delta t^\text{obs} &= t^\text{corr}_\text{ec-tr}(M,N) + t^\text{corr}_\text{tr-ec}(0,-7) \\
&= \bar{t}_m(M) - t_m(N) - (M-N) P + 5.85445768\, \mathrm{d}.\nonumber
\end{align}
Using the TESS transit ($N=33$) and MIRI eclipse ($M=50$), we obtain 
\begin{equation}\label{TESSMIRI}
\Delta t^\text{obs} = 507.9 \pm 157.0 \, \mathrm{s},
\end{equation}
which differs from zero by $3.24 \sigma$, indicating that the purely Keplerian, zero-precession model is inconsistent with our observations.

On the other hand, 
using the GR-only predictions calculated in Section~\ref{sec:theoretical predictions} from the updated system parameters (Table~\ref{tab:planet_parameters}), the expected timing shifts for the TESS transit at orbit $N=33$ and the JWST/NIRSpec eclipse at orbit $M=42$, 
and for the TESS transit at orbit $N=33$ and the JWST/MIRI eclipse at orbit $M=50$, the corresponding predictions are (see Table \ref{tab:gr_only_predictions})
\begin{subequations}
\begin{align}
    \Delta t^{\rm GR} &= 185.07~\text{s}\quad\text{(TESS-JWST/NIRSpec)},\\
    \Delta t^{\rm GR} &= 187.67~\text{s}\quad\text{(TESS-JWST/MIRI)}.
\end{align}
\end{subequations}
These predictions represent an important improvement but the observed timing shift~\eqref{TESSMIRI} still exceeds the GR-only prediction, for instance by 320.2\,s in the case of the TESS-JWST/MIRI timing combination $(M,N)=(50,33)$, corresponding to a discrepancy of $2.04\sigma$ when considering the observational timing uncertainty alone.

As a consistency check, we also calculated the timing combinations using the two NIRSpec eclipse measurements at $(M,N)=(42,33)$. The resulting values, presented alongside the MIRI measurement in Table~\ref{tab:gr_timing}, are consistent within their uncertainties. We adopt the MIRI measurement for the following comparison, as it provides the smallest timing uncertainty and the longest observational baseline.

\begin{table*}
    \centering
    \caption{Observed relativistic timing combinations compared with the updated purely GR (no tides) predictions. The quoted significance relative to the GR prediction considers the observational timing uncertainty only.}
    \label{tab:gr_timing}
    \begin{tabular}{lcccc}
        \hline \hline
        Dataset &
        $\Delta t^{\mathrm{obs}}$ (s) &
        $\Delta t^{\mathrm{GR}}$ (s) &
        $\sigma$ from zero &
        $\sigma$ from GR \\
        \hline
        NIRSpec NRS1 & $568.2 \pm 175.2$ & 185.07 & 3.24 & 2.19 \\
        NIRSpec NRS2 & $474.1 \pm 181.0$ & 185.07 & 2.62 & 1.60 \\
        MIRI         & $507.9 \pm 157.0$ & 187.67 & 3.24 & 2.04 \\
        \hline
    \end{tabular}
\end{table*}

Using the system parameters from the global fit in Table \ref{tab:planet_parameters}, the predicted GR apsidal precession rate is
\begin{align}\label{GR-210}
\dot{\omega}^{\rm GR}
\equiv n\,k^{\rm GR} &= 0.0583^{+0.0018}_{-0.0017}~
\mathrm{deg~century^{-1}}\nonumber\\
&= 209.9^{+6.5}_{-6.1}~
\mathrm{arcsec~century^{-1}},
\end{align}
where $n=2\pi/P$ and $k^{\rm GR}$ is the relativistic parameter~\eqref{kGR}. 
To facilitate comparison with both the predicted GR apsidal precession rate and the rate directly fitted by \citet{beichman2026eclipse}, we convert our observed timing shift into an effective precession rate, $\dot{\omega}_{\rm eff}$,
 as the rate that would reproduce our observed timing shift,
assuming that the timing observable depends linearly on the apsidal precession rate.\footnote{This can be justified by the fact that in the timing formula for $\delta t$ given by \eqref{deltat}, the only secular-growing term (besides many periodic terms) is the precession rate $k$ which enters linearly in the timing formula through $\delta\psi$ given by \eqref{deltapsi} and then linearly through the variation of $Y(\psi)$, see the secular term $\propto k\,v$ in Eq. \eqref{Y}.} This linear approximation is appropriate for the small accumulated apsidal advance expected over the observational baseline.
The conversion factor derived from the GR calculation is
\begin{equation}
\frac{\dot{\omega}^{\mathrm{GR}}} {\Delta t^{\mathrm{GR}}} = 1.1185~ \frac{\mathrm{arcsec~century^{-1}}}{\mathrm{s}}.
\end{equation}
Applying this conversion to the measured MIRI timing combination gives 
\begin{align}
\dot{\omega}_{\mathrm{eff}} = \frac{\dot{\omega}^{\mathrm{GR}}} {\Delta t^{\mathrm{GR}}}\Delta t^{\mathrm{obs}}
&= 568.0^{+176.5}_{-176.4}~\mathrm{arcsec~century^{-1}} \nonumber\\ &= 0.1578^{+0.0490}_{-0.0490}~ \mathrm{deg~century^{-1}}.
\end{align}

The uncertainty includes both the observational uncertainty in $\Delta t^{\mathrm{obs}}$ and the uncertainty in the predicted GR precession rate, added in quadrature. The uncertainty is dominated by the timing measurement, which contributes approximately $175.6$\,arcsec\,century$^{-1}$, compared with $^{+17.6}_{-16.5}$\,arcsec\,century$^{-1}$ from the uncertainty in $\dot{\omega}^{\rm GR}$. 


 The inferred effective precession rate exceeds
the GR prediction \eqref{GR-210} by
$358.1$\,arcsec\,century$^{-1}$,
corresponding to the same $2.04\sigma$ discrepancy found in timing space.
However, this difference is inferred from the observed timing shift rather than from a direct measurement of an additional precession component.

For comparison, \citet{beichman2026eclipse} obtained $\dot{\omega}=302^{+294}_{-289}$\,arcsec\,century$^{-1}$. Our inferred effective rate, $\dot{\omega}_{\rm eff}=568.0^{+176.5}_{-176.4}$ \,arcsec\,century$^{-1}$, is higher, although the two estimates are compatible within their quoted uncertainties.

Unlike our approach, \citet{beichman2026eclipse} fitted a secular precession rate within a global orbital model combining radial-velocity measurements with individually determined transit and eclipse mid-times, rather than the underlying photometric light curves. In contrast, we use a new global fit to the photometric and radial-velocity data to establish a Keplerian reference solution, followed by independent event-timing measurements and application of the perturbative formalism of \citet{blanchet2019detecting} to construct a differential transit--eclipse timing observable.

Our approach therefore provides an independent and complementary constraint on apsidal precession. While the inferred effective precession rate is consistent with that of \citet{beichman2026eclipse} within uncertainties, the different methodologies and statistical treatments can lead to different constraints on the precession rate.

\section{Discussion}
\label{sec:discussion}

The timing combinations derived from the NIRSpec and MIRI eclipses are consistently positive, with the MIRI measurement differing from zero at $3.24 \sigma$. However, the measured shift of $507.9 \pm 157.0$\,s, see Eq.~\eqref{TESSMIRI}, exceeds the GR-only prediction of $187.67$ s by $2.04 \sigma$. Although the observed timing shift provides evidence for apsidal precession, its nominal excess over the GR-only prediction motivates an investigation of other possible contributions.

\subsection{Tidal contribution}
\label{subsec:tidal_contribution}

Tidal deformation provides an additional contribution to the apsidal precession and timings, the magnitude of which depends on the planet's and star's internal structure through the Love number $k_p^{(2)}$ and $k_\star^{(2)}$, where the superscript $(2)$ refers to the quadrupolar deformation~\citep{Love11}.\footnote{We adopt the Love-number convention, where $k^{(2)}$ differs by a factor 2 from the so-called apsidal motion constant,  $k^{(2)}=2k_\text{aps}^{(2)}$ \citep{ragozzine2009probing}.} The value of $k^{(2)}$ depends on the internal density distribution, with increasing central concentration producing smaller Love numbers \citep{kramm2011degeneracy}. For reference, an $n=1$ polytrope has $k^{(2)} \simeq 0.52$, while recent measurements for Hot Jupiters span approximately $k^{(2)} \sim 0.2-0.6$ \citep{van2025retrieving}. We therefore explore $k_p^{(2)} \sim 0.2-0.6$ as a plausible range for HD 80606 b. 
This range 
is illustrative rather than a specific constraint on the internal structure of HD~80606~b.
 Given its relatively high mass ($M_p\simeq 4\,M_{\rm Jup}$) with a radius of $R_p\simeq 1.01\,R_{\rm Jup}$, the planet is substantially denser than many inflated hot Jupiters for which Love-number
estimates have been obtained.
The quadrupolar Love number depends on the internal density distribution, and dedicated interior models would be required to establish a more restrictive
range for this planet.

For the star we may consider $k_\star^{(2)}$ between $0$ and $0.2$. In a binary system the relevant combination of Love numbers is the tidal polarizability parameter given by Eq.~\eqref{mu+2}, \textit{i.e.} 
\begin{subequations}
\begin{align}
	\mu_+^{(2)} \equiv \frac{1}{3}\biggl[ \frac{M_\star}{M_p} \,k_p^{(2)} R_p^5 + \frac{M_p}{M_\star} \,k_\star^{(2)} R_\star^5\biggr],
\end{align}
and the tidal contribution to apsidal precession is \citep{jordan2008observability, Fabrycky}
\begin{equation}
\dot{\omega}^\mathrm{tide}
	= \frac{45}{8} \frac{n(8+12e^2+e^4)}{a^5(1-e^2)^5}
	\,\mu_+^{(2)}.
\end{equation}
\end{subequations}
See Eqs.~\eqref{parameterstide} for the complete tidal contributions to the Newtonian binary motion. With the Love numbers $k_p^{(2)}\simeq 0.6$ and $k_\star^{(2)}\simeq 0.1$ we find that the tidal apsidal precession is $\dot{\omega}^\text{tide}\simeq 97\,\mathrm{arcsec~century^{-1}}$, of which
$\simeq 77$ is due to the planet and $\simeq 20$ is due to the star. The tidal effect is therefore non-negligible although smaller than the GR precession~\eqref{GR-210}. 

As discussed in Section~\ref{sec:theoretical predictions}, the perturbative timing formalism allows the tidal contribution to be calculated alongside the GR prediction. Here, we isolate the tidal contribution, excluding rotational deformation, which is considered separately in Section~6.2.

\begin{table*}
\centering
\caption{Comparison of observed transit--eclipse
timing shifts with theoretical predictions including
general relativity and Newtonian quadrupolar tidal
deformation: $\Delta t^{\rm pred} = \Delta t^{\rm GR} + \Delta t^{\rm tide}$. Unlike Table~\ref{tab:gr_tides_kstar}, rotational contributions are excluded from these predictions. For the tidal models, the stellar Love
number is fixed at $k_\star^{(2)}=0.2$, while
$k_p^{(2)}$ is varied. The GR-only predictions
correspond to $k_p^{(2)}=k_\star^{(2)}=0$.
Differences are calculated as
$\Delta t^{\rm obs}-\Delta t^{\rm pred}$.
Significances include observational timing and
orbital-period uncertainties only, excluding
uncertainties in the theoretical predictions.}
\label{tab:timing_tidal_comparison}
\begin{tabular}{llcccc}
\hline\hline
Dataset & $k_p^{(2)}$ & $\Delta t^{\rm obs}$ (s) & $\Delta t^{\rm pred}$ (s) & Difference (s) & $\sigma$ \\
\hline
MIRI & GR only & $507.9\pm157.0$ & 187.67 & 320.23 & 2.04 \\
& 0.2 & & 243.70 & 264.20 & 1.68 \\
& 0.4 & & 265.49 & 242.41 & 1.54 \\
& 0.6 & & 287.28 & 220.62 & 1.41 \\
\hline
NRS1 & GR only & $568.2\pm175.2$ & 185.07 & 383.13 & 2.19 \\
& 0.2 & & 241.17 & 327.03 & 1.87 \\
& 0.4 & & 262.98 & 305.22 & 1.74 \\
& 0.6 & & 284.80 & 283.40 & 1.62 \\
\hline
NRS2 & GR only & $474.1\pm181.0$ & 185.07 & 289.03 & 1.60 \\
& 0.2 & & 241.17 & 232.93 & 1.29 \\
& 0.4 & & 262.98 & 211.12 & 1.17 \\
& 0.6 & & 284.80 & 189.30 & 1.05 \\
\hline
\end{tabular}
\end{table*}

Using the updated orbital parameters and the perturbative timing formalism introduced in Section~4, we calculate the combined GR and tidal predictions for the TESS--NIRSpec and TESS--MIRI timing combinations. Table~\ref{tab:timing_tidal_comparison} compares these predictions with the measured timing shifts for a fixed stellar Love number of $k_\star^{(2)}=0.2$ and planetary Love numbers of $k_p^{(2)}=0.2$--$0.6$. Unlike the predictions in Table~7, these calculations exclude rotational deformation.

For the TESS--MIRI combination, including tidal deformation increases the predicted timing shift from $187.67$\,s for GR alone to $243.70$--$287.28$\,s. The observed shift of $507.9\pm 157.0$\,s therefore exceeds the combined GR and tidal predictions by $220.6$--$264.2$\,s, corresponding to discrepancies of $1.41$--$1.68\sigma$, compared with $2.04\,\sigma$ for GR alone. The NIRSpec measurements show a similar reduction, with discrepancies of $1.62$--$1.87\sigma$ for NRS1 and $1.05$--$1.29\sigma$ for NRS2.

Thus, although the central measured timing shifts remain larger than the theoretical predictions, including tidal deformation improves agreement with the observations. The remaining differences are below $2\sigma$ for all three datasets and all explored tidal models, indicating that the current observations are consistent with the combined contributions of GR and tides, without requiring an additional mechanism. These comparisons consider observational timing and orbital-period uncertainties only, excluding uncertainties in the theoretical predictions.

\subsection{Rotational contribution}
\label{subsec:rotational_contribution}

Rotational deformation of the planet and host star can also contribute to apsidal precession. Assuming zero obliquity, the rotational contribution from each body can be expressed as \citep{ragozzine2009probing}: 
\begin{subequations}
\label{eq: rotational_aps_precession}
\begin{equation}
\dot{\omega}^\text{rot} = \frac{n}{2a^2(1-e^2)^2} \lambda_+^{(2)},
\end{equation}
where the effective combination $\lambda_+^{(2)}$ of the planet's and star's rotation $\Omega_p$ and $\Omega_\star$ is given by~\eqref{lambda+2}, \textit{i.e.}
\begin{align}
\lambda_+^{(2)} \equiv k_p^{(2)} \frac{R_p^5\Omega_p^2}{G M_p} + k_\star^{(2)} \frac{R_\star^5\Omega_\star^2}{G M_\star} ,
\end{align}
\end{subequations}
See Eqs.~\eqref{parametersrot} for the complete rotational contributions. 

For HD 80606 b, \citet{wit2016direct} inferred a planetary rotation period of $93^{+85}_{-35}$\,h, while pseudo-synchronous rotation corresponds to approximately $40.5$\,h. Adopting the range $k_p^{(2)}\sim 0.2$--$0.6$ gives $\dot{\omega}_p^{\rm rot} \sim 0.70$--$2.09$\,arcsec\,century$^{-1}$ for the inferred rotation period, increasing to $3.67$--$11.01$\,arcsec\,century$^{-1}$ for pseudo-synchronous rotation. The stellar rotational contribution is smaller still and does not alter this conclusion. These contributions are small compared with the GR prediction of $209.9^{+6.5}_{-6.1}$\,arcsec\,century$^{-1}$ and therefore cannot account for the nominal excess in the timing-equivalent precession rate.

HD 80606 b exhibits significant spin--orbit misalignment \citep{Hebrard2010HD80606b}, which can modify the rotationally induced apsidal precession and introduce nodal precession. The rotational contributions estimated above correspond to timing shifts of approximately $0.6$--$1.9$\,s for the inferred planetary rotation period, or $3.3$--$9.8$\,s for pseudo-synchronous rotation. Even allowing for geometrical corrections associated with spin--orbit misalignment, these effects are expected to remain small compared with the current timing uncertainty of approximately $157$\,s.

As an illustrative upper-limit scenario, we further investigated whether a sufficiently rapidly rotating planet could account for the observed excess precession. Assuming that the entire nominal excess of $358.1$\,arcsec\,century$^{-1}$ above the GR prediction arises from rotational deformation, we rearranged Eqs.~(17) to calculate the required planetary rotation period as a function of the potential Love number $k_p^{(2)}$. For the adopted $k_p^{(2)}=0.2$--$0.6$, the required rotation periods range from approximately $4.1$ to $7.1$\,h, as illustrated in Fig.~6. These periods are substantially shorter than both the phase-curve estimate of $93^{+85}_{-35}$\,h \citep{wit2016direct} and the pseudo-synchronous rotation period of approximately $40.5$\,h. Thus, while sufficiently rapid planetary rotation could in principle account for the nominal excess, it would be inconsistent with the existing phase-curve constraint. The stellar rotational contribution is smaller still, given the host star's rotation period of approximately 40\,d (Table~\ref{tab:exofast_stellar}).

\begin{figure}
    \centering
    \includegraphics[width=\columnwidth]{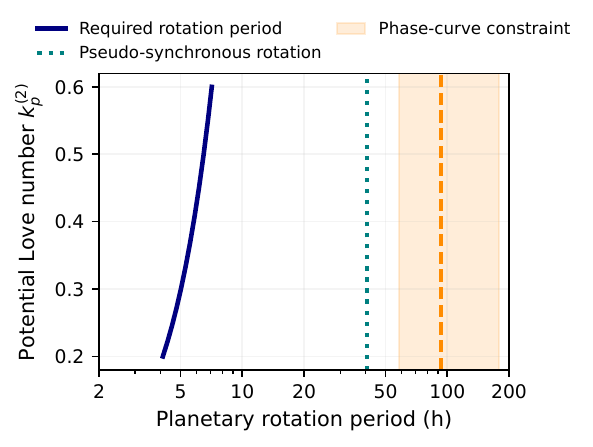}
    \caption{Planetary rotation period required to reproduce the nominal excess apsidal (effective) precession of $358.1$\,arcsec\,century$^{-1}$ above the GR-only prediction, as a function of the planetary potential Love number $k_p^{(2)}$. The dotted line indicates the pseudo-synchronous rotation period of approximately $40.5$\,h, while the dashed line and shaded region show the phase-curve estimate of $93^{+85}_{-35}$\,h from \citet{wit2016direct}. The calculations assume zero obliquity and attribute the entire nominal excess above GR to planetary rotational deformation.}
    \label{fig:rotation_love}
\end{figure}

\subsection{Influence of HD80607}

HD~80607 is unlikely to contribute appreciably to the timing evolution considered here. The companion lies at an angular separation of $\sim20.5''$ from HD~80606, while the fitted distance to the system is $\simeq65.9$~pc, corresponding to a projected physical separation of approximately $1.35\times10^3$~AU. At such a large separation, the secular perturbation exerted by HD~80607 on the $\sim0.45$~AU orbit of HD~80606~b is expected to operate on timescales far longer than the $\sim$ decade baseline of the present timing measurements. Although the wide stellar companion may have played an important role in the long-term dynamical evolution of the system, for example by modifying the planetary orbit over secular timescales, its instantaneous contribution to the apsidal precession over the 2007--2025 interval should be negligible compared with the relativistic and tidal terms. It is therefore unlikely that HD~80607 can account for the observed few-hundred-second excess in the eclipse--transit timing residuals.

\subsection{A satellite companion to HD~80606~b?}

A potential alternative explanation for the nominal timing excess over the GR-only prediction is the presence of a satellite orbiting HD~80606~b. However, the extreme eccentricity of the planetary orbit places severe constraints on the long-term stability of such a companion. Using the updated system parameters from our global fit in Table~\ref{tab:planet_parameters},
we obtain a periastron distance of $r_{\rm peri}\simeq0.0313$~AU. The corresponding periastron Hill radius is
\begin{equation}
r_{\rm H,peri}=a(1-e)
\left(\frac{M_p}{3M_\star}\right)^{1/3}
\simeq\,5.1\times10^5~{\rm km},
\end{equation}
or only $\sim7.1\,R_p$. The region available for stable satellite orbits is therefore severely restricted. For comparison, illustrative Roche limits are approximately $2.9\,R_p$ for a rocky satellite and $4.2\,R_p$ for an icy satellite, although their exact values depend on the assumed densities and internal structures. Adopting an approximate stability criterion of $a_m\lesssim0.3\,r_{\rm H,peri}$ gives $a_m\lesssim2.1\,R_p$, already interior to these Roche limits. This suggests that the region in which a conventional long-lived moon could survive is extremely limited. However, the adopted stability criterion is only approximate, particularly for such an eccentric planetary orbit, and a definitive assessment would require dedicated dynamical simulations.

A sufficiently massive satellite could also induce transit and eclipse timing variations through the reflex motion of the planet around the planet--satellite barycentre. The characteristic timing amplitude can be approximated by
\begin{equation}
\Delta T \sim \frac{a_m}{v_\perp}\frac{M_m}{M_p},
\end{equation}
where $a_m$ and $M_m$ are the satellite semi-major axis and mass, respectively, and $v_\perp$ is the projected velocity of the planet during the event. 
Adopting $v_\perp\simeq24.4$~km~s$^{-1}$, a $300$~s timing amplitude would require satellite masses of approximately $66$, $44$, and $26\,M_\oplus$ at separations of $2$, $3$, and $5\,R_p$, respectively. 

These estimates assume a favourable orbital projection and imply unusually massive satellites, comparable to Neptune-like planets. Moreover, the larger orbital separations exceed the adopted stability limit.
Satellite-induced timing variations depend on the satellite's orbital phase and do not necessarily mimic secular apsidal precession. Since our timing observable combines transit and eclipse measurements from different epochs, the estimated satellite masses should be regarded as order-of-magnitude estimates rather than direct constraints.

Although a satellite cannot be ruled out without dedicated simulations, the restricted stable region and large masses required make it an unlikely explanation for the nominal excess over the GR-only prediction.

\subsection{Additional planetary companion}
\label{companion}
An additional planetary companion could provide a
further contribution to the observed timing evolution
through gravitational perturbations. To investigate
this possibility, we performed three RV-only \exofast{}
fits comparing no long-term RV trend to both a linear and a quadratic trend. The analysis included the same RVs used in the full global fit, \textit{i.e.} 270
measurements from seven instruments, spanning
approximately 22 years (1999--2021; see Table \ref{tab:rv_observations}).
Neither fit incorporating long-term trends yielded a statistically significant trend,
with the fitted coefficients consistent with zero
within approximately $1$--$2\sigma$. The BIC statistic for each fit was 2029.7, 2035.4 and 2041.3 for the flat, linear and quadratic trends, respectively, indicating the RVs are not better characterised with an increasing number of fit parameters.
The residuals of the combined multi-instrument
RV dataset exhibit an RMS scatter of approximately
$4$\,m\,s$^{-1}$, after subtraction of the fitted
orbital model and instrumental offsets.
Although this scatter does not constitute a formal
detection threshold, it suggests that additional
companions producing RV semi-amplitudes of
a few m\,s$^{-1}$ could remain undetected.
An exploratory periodogram analysis of these
residuals did not identify any obvious additional
periodic signal. This preliminary examination provides no indication of an additional planetary companion contributing
to the RV data.

Determining the range of parameters for a potentially unseen companion capable of producing the nominal precession excess can be approached analytically in a similar manner to  \cite{Heyl_2007} and \cite{jordan2008observability}. Here, we use a standard, double orbit-averaged, quadrupole-order secular approximation of the three-body problem, assuming coplanar orbits with $ a_b \ll a_c $. The resulting quadrupole-order precession rate is
\begin{equation}\label{eq:qauand}
    \dot{\omega}_{\rm{quad}} = \frac{3}{4}n_b \frac{M_c}{M_*} \left(\frac{a_b}{a_c} \right)^3 \frac{(1-e_b^2)^{1/2}}{(1-e_c^2)^{3/2}},
\end{equation}
where the parameters for the inner planet and the hypothetical companion are denoted by the subscripts $b$ and $c$, respectively, and $n_b = 2\pi/P_b$ is the mean motion of the inner planet.\\
However, truncating the secular disturbing function at quadrupole order neglects the azimuthal asymmetry of an eccentric outer perturber's orbit, as $e_c$ enters Eq.~\eqref{eq:qauand} solely through the orbit-averaged factor $(1-e_c^2)^{-3/2}$. Because HD 80606 b is on a highly eccentric orbit ($e_b \approx 0.93$), any eccentric outer companion ($e_c > 0$) exerts an asymmetric gravitational torque that a symmetric quadrupole ring approximation will not capture well. To account for this lopsided potential and evaluate a wide range of plausible precession rates, we extend the expansion to octupole order \citep[e.g.,][]{Ford_2000,Lee_2003, Naoz_2013}:
\begin{equation}
    \dot{\omega}_{\rm{oct}} = -\frac{15}{64} n_b \frac{M_c}{M_*} \left(\frac{a_b}{a_c}\right)^4 \frac{(1-e_b^2)^{1/2}(4 + 9e_b^2) e_c \cos\Delta\varpi}{e_b (1-e_c^2)^{5/2}},
\end{equation}
where $\Delta\varpi = \varpi_b - \varpi_c$ is the difference in the longitudes of the pericenter between the two orbits. The total perturbation precession rate is then $\dot{\omega}_{\rm{pert}} = \dot{\omega}_{\rm{quad}} + \dot{\omega}_{\rm{oct}}$.
For $e_b = 0.93$, the eccentricity factor is evaluated as $(4 + 9e_b^2)/e_b \approx 12.67$, which means that even for moderate outer eccentricities ($e_c \lesssim 0.3$) and semi-major axis ratios $a_b/a_c \gtrsim 0.05$, the octupole term alters the predicted precession rate by tens of percent. From this analysis, multiple combinations of companion planet parameters can produce a precession rate compatible with the nominal excess. For example, a 0.3 $M_{\rm{Jup}}$ companion at 2--3 AU with an eccentricity between 0.05 and 0.6 would be sufficient, whereas a more massive 1 $M_{\rm{Jup}}$ companion with the same eccentricity range would require a semi-major axis between 3 and 5 AU.

As a further check of plausible planetary companion parameters, we performed preliminary $N$-body simulations using \textsc{rebound} \citep{Rein2012} with the \textsc{IAS15} integrator \citep{Rein_2015}. \textsc{IAS15} is a 15th-order Gauss--Radau integrator belonging to the family of high-order Runge--Kutta methods, which have been shown to outperform the orbital precision of lower order methods and achieve stability over tens of millions of years \citep[e.g.][]{Walters_2022}. Importantly, its adaptive time-stepping scheme refines the step size during close approaches, allowing us to accurately resolve the rapid orbital motion of HD 80606 b near periastron while maintaining long-term numerical stability. We explored 1,400 companion configurations over 10,000 orbits, varying the semi-major axis ($1-5$ AU), inclination ($60^{\circ} - 90^{\circ}$), and longitude of ascending node ($0^{\circ} - 90^{\circ}$). Of these, 334 sub-Jovian configurations reproduced the nominal timing excess. Relaxing the coplanar and orbit-averaged assumptions by including mutual orbital inclinations and capturing short-timescale phase-dependent perturbations allowed substantially lower-mass \(M_c \approx 2.8\text{--}39.2\,M_\oplus\) companions spanning semi-major axes of \(a_c = 1.5\text{--}3.0\,\mathrm{AU}\) to match the timing excess. These configurations yield predicted stellar radial-velocity semi-amplitudes of approximately $0.16\text{--}2.08\,\mathrm{m\,s^{-1}}$ and could readily evade RV detection. More detailed analysis using $N$-body simulations is deferred to future work.

\section{Conclusion}
\label{sec:conclusion}

\begin{table*}[t]
\caption{Summary of the observed timing shift and the possible physical contributions to apsidal precession in HD~80606~b. The nominal excess refers to the difference between the observed timing shift and the GR-only prediction.}
\label{tab:precession_mechanisms}
\centering
\begingroup
\renewcommand{\arraystretch}{1.6}
\setlength{\tabcolsep}{9pt}

\begin{tabular}{p{2.8cm}p{6.7cm}p{5.9cm}}
\hline\hline
Mechanism & Measured or predicted effect & Interpretation \\
\hline
\noalign{\vskip 3pt}
\shortstack[c]{Obs. timing shift \\
($\Delta t^\text{obs}$)}
& $\Delta t^{\rm obs}=507.9\pm157.0$~s,
derived from the TESS transit and JWST/MIRI eclipse timings,
relative to the reference Spitzer observations and
the purely Newtonian orbital prediction.
& Measured relative to the purely Newtonian prediction,
corresponding to a $3.24\sigma$ departure from zero. \\
\hline
\noalign{\vskip 3pt}
\shortstack[c]{ General relativity  \\
($\Delta t^{\text{GR}}$)}
& Predicted apsidal precession of
$209.9^{+6.5}_{-6.1}$~arcsec~century$^{-1}$,
corresponding to a timing shift of $\Delta t^{\text{GR}} \sim 187.7$~s.
& Predicts a non-zero timing shift that is smaller
than the observed value by $2.04\sigma$. \\

\shortstack[c]{Tidal deformation \\
($\Delta t^{\text{tide}}$)}
& Tidal timing shifts of approximately
$\Delta t^{\rm tide}\sim56$--$100$\,s for
$k_\star^{(2)}=0.2$ and $k_p^{(2)}=0.2$--$0.6$,
excluding rotational deformation. The total
GR-plus-tidal timing shift is
$\Delta t^{\rm GR}+\Delta t^{\rm tide}
\simeq244$--$287$\,s over the explored
Love-number range.
& Contributes to apsidal precession and increases
the predicted timing shift. When combined with GR,
tidal deformation reduces the discrepancy with the
observations to $1.41$--$1.68\sigma$ for the
TESS--MIRI combination. \\

\shortstack[c]{Planetary rotation\\
($\Delta t^{\rm rot}$)}
&  For the measured rotation period of
$93^{+85}_{-35}$\,h, the approximate rotational
timing contribution is $0.6$--$1.9$\,s,
extending to $0.2$--$4.8$\,s over its uncertainty
range. Pseudo-synchronous rotation ($40.5$\,h)
would produce approximately $3.3$--$9.8$\,s,
for $k_p^{(2)}=0.2$--$0.6$.
&
Planetary rotation contributes negligibly at the
expected rotation rate. Explaining the nominal
excess above GR through planetary rotation alone
would require periods of $4.1$--$7.1$\,h,
substantially shorter than the phase-curve
estimate of $93^{+85}_{-35}$\,h. \\

\shortstack[c]{  Stellar spin--orbit\\misalignment}
&  For $P_\star\simeq40$\,d and
$k_\star^{(2)}=0.2$, the stellar rotational
quadrupole gives a characteristic apsidal
contribution of $\sim2.7$ arcsec century$^{-1}$.
The associated nodal precession is
$\lesssim2.7$ arcsec century$^{-1}$.
& Stellar rotational apsidal contribution
corresponds to a timing shift of a few seconds.
A precise nodal timing correction requires
the three-dimensional spin--orbit geometry. \\ 

Stellar companion HD~80607
& Secular perturbations operate on timescales
much longer than the observational baseline.
& Its contribution to apsidal precession is expected
to be negligible over the timescale considered here. \\

Satellite companion
& The small periastron Hill sphere severely restricts
the region available for stable satellite orbits.
Producing a timing amplitude of several hundred seconds
would require an unusually massive companion.
& A long-lived satellite appears dynamically disfavoured,
although dedicated stability and timing simulations
would be needed for a definitive assessment. \\

Additional planet
&  RV measurements do not show clear evidence for a companion planet. Analytically predicted companion masses $\gtrsim 0.3 \ M_{\rm{Jup}}$ and semi-major axes (2-5 AU) are larger than predicted by $N$-body models which additionally yield small RV amplitudes of $0.16\text{--}2.08\,\mathrm{m\,s^{-1}}$.
&  Many planetary companion properties are capable of producing the nominal timing excess and some yield small RV amplitudes which could make detection difficult. \\

\\

\hline
\end{tabular}
\endgroup
\end{table*}

We have revisited the search for apsidal precession in the highly eccentric exoplanet HD 80606 b using transit and secondary-eclipse observations spanning approximately 17 years, together with more than two decades of radial-velocity measurements. By combining re-reduced Spitzer photometry, re-extracted and detrended TESS observations, a partial ground-based transit, and recent JWST/NIRSpec and MIRI observations in a global fit with the radial-velocity data, we refined the physical and orbital parameters of the system and revisited the theoretical predictions of \citet{blanchet2019detecting}.

Comparing the period-corrected transit--eclipse separation from the earlier Spitzer observations with that obtained from the later TESS transit and JWST/MIRI eclipse, we measured a timing shift of $ \Delta t^\text{obs} =507.9 \pm 157.0$ s relative to the purely Newtonian prediction, corresponding to a $3.24 \sigma$ departure from zero. This suggests measurable apsidal evolution in HD 80606~b. The observed shift exceeds the GR-only prediction of approximately 188 s by $2.04 \sigma$. Including Newtonian tidal deformation increases the predicted timing shift to approximately 244--287 s for the explored Love-number combinations, reducing the discrepancy to $1.41$--$1.68 \sigma$. 

These results provide the first observational indication of apsidal precession in an exoplanetary system where general relativity is expected to be the dominant contribution. Unlike previously reported cases of exoplanetary apsidal precession, which are dominated by classical effects or remain incompletely explained, HD 80606 b offers an opportunity to probe relativistic orbital evolution through differential transit--eclipse timing measurements. Although the present observations do not independently isolate the relativistic contribution, improved timing precision and tighter constraints on classical effects will enable more stringent tests of the relativistic predictions.

We also investigated possible additional contributions to the timing evolution. Planetary rotational deformation, the wide stellar companion HD 80607, and a hypothetical satellite are unlikely to account for the nominal excess over the GR-only prediction. 

{ An additional planetary companion remains a possible
contribution to the nominal timing excess, although
our radial-velocity analysis does not reveal a
statistically significant additional signal.
Analytical estimates and exploratory} N-body simulations do not uniquely
constrain the orbital configuration of such a companion.
Results are summarized in Table  \ref{tab:precession_mechanisms}.

Future observations with JWST and Ariel will be crucial for refining the transit and eclipse timings of HD 80606 b and extending the observational baseline. Their high photometric precision will improve constraints on apsidal precession and help distinguish between the contributions of general relativity, tidal deformation, and possible additional dynamical perturbations. Continued monitoring will also provide valuable opportunities to further characterise the orbital and atmospheric properties of this extreme exoplanet.

\begin{acknowledgements}
This work originated as an Honours research project undertaken by S.-L. Buelens at the University of Tasmania and was subsequently developed into the present study. S.-L. Buelens acknowledges the hospitality of the
Institut d'Astrophysique de Paris (IAP), where part of this work was carried out. J.-P. Beaulieu warmly acknowledges Sean Carey (IPAC) for his friendship and for numerous insightful exchanges about Spitzer and infrared data over the past two decades. L. Blanchet thanks Guillaume H\'ebrard and Quentin Henry for informative discussions.

This work was supported by the University of Tasmania through the endowed Warren Chair in Astronomy and by the Australian Government through the Australian Research Council Discovery Project grants DP200101909 and DP240101842. Research in France was supported by the SPACE-MLENS project (ANR-24-CE31-3263) and CNES.

This research made use of Lightkurve, a Python package for Kepler and TESS data analysis \citep{Lightkurve:2018}, and Astropy, a community-developed core Python package and ecosystem of tools and resources for astronomy \citep{astropy:2013, astropy:2018, astropy:2022}.

This work makes use of archival data products from the
Spitzer Space Telescope, which was operated by the
Jet Propulsion Laboratory, California Institute of Technology,
under a contract with NASA. The Spitzer observations
were obtained under programmes 40386 (PI: G. Laughlin)
and 540 (PI: G. H\'ebrard).

This work also makes use of archival data products from the
NASA/ESA/CSA James Webb Space Telescope, retrieved from the
Mikulski Archive for Space Telescopes (MAST) at the
Space Telescope Science Institute, which is operated by
the Association of Universities for Research in Astronomy,
Inc., under NASA contract NAS5-03127.

The JWST observations were obtained under programmes
GO-2488 (PI: J. Sikora) and GO-2008 (PI: T. Kataria).
The NIRSpec observations are available from MAST
at \url{https://doi.org/10.17909/yb30-hj11}.    
\end{acknowledgements}


\bibliography{HD80606b_GR.bib}

\begin{appendix}
 \nolinenumbers

\section{TESS photometry}
\label{app:tess}
The TESS observations were accessed through the \textit{Lightkurve} Python package \citep{Lightkurve:2018} from the Mikulski Archive for Space Telescopes (MAST). We considered the light-curve products processed by the Science Processing Operations Center (SPOC; \citealt{Jenkins:2016}) pipeline. This pipeline calibrates the raw observations, performs aperture photometry, corrects instrumental effects and contamination, and conducts transit-search diagnostics. Its Pre-search Data Conditioning Simple Aperture Photometry (PDCSAP) products are designed to remove most instrumental variability and long-term trends \citep{Stumpe:2012,Stumpe:2014,Smith:2012}.

The nearby companion HD 80607 lies approximately 20.5~arcsec from HD 80606, corresponding to roughly one TESS pixel, so the two stars are blended in TESS images. The default SPOC photometric aperture did not adequately encompass HD 80606 while still including significant flux from its companion. We therefore used a custom, wider aperture, selecting pixels with a median flux above $50\,\mathrm{e}^{-}\,\mathrm{s}^{-1}$ to include both stars. Contamination by HD 80607 was subsequently accounted for through a dilution parameter in the global fit (Section~\ref{sec:updated_system_params}).

Because the transit midpoint is central to our timing analysis, we assessed whether the detrending procedure could bias the recovered transit parameters. Before extracting the final detrended light curve, we performed transit-injection and recovery tests away from the observed transit, using \texttt{batman} \citep{batman:2015} to model the injected events. We compared the Savitzky--Golay filter available in \textit{Lightkurve} with the spline-based \texttt{keplerspline}\footnote{\url{https://github.com/avanderburg/keplersplinev2}} method \citep{Vanderburg:2014}. For both methods, we explored 30 evenly spaced window lengths or knot spacings between 0.7 and 4.0~d and assessed the recovery of the transit depth, duration $T_{14}$, and midpoint $T_0$.

The Savitzky--Golay filter marginally improved recovery of the transit depth and duration, whereas the spline method provided more reliable recovery of the transit midpoint. We therefore adopted \texttt{keplerspline} with a knot spacing of 1.27~d (approximately $2.5\,T_{14}$) to remove the remaining stellar variability and instrumental trends from the out-of-transit light curve. The resulting detrended photometry was used in the global fit.

\section{Spitzer photometry}
\label{app:spitzer}

For both observations, we started from the Basic Calibrated Data (BCD) products. 
For the 4.5 $\mu$m transit, each BCD cube contains 64 consecutive $32 \times 32$ pixel subarray frames. The frames within each cube were median-combined to produce one image per cube, with the timestamp assigned to the midpoint of the 64-frame sequence. The position of HD 80606 was measured in each stacked image using a two-dimensional Gaussian centroid within a $9 \times 9$ pixel region. Aperture photometry was performed using a 3-pixel radius, with the background estimated from a 6-10 pixel annulus using a $3 \sigma$-clipped median. Centroid outliers and subsequent photometric residual outliers were rejected using a $3\sigma$ robust-MAD criterion, with  the same criterion subsequently applied to the photometric residuals. 

To account for the IRAC intrapixel sensitivity \citep{Reach2005} we followed \citet{Beaulieu2008}. The transit and position-dependent sensitivity were fitted simultaneously as 
\begin{equation}
F_{\rm model}(t)
=
F_{\rm tr}(t)\, S[x(t),y(t)],
\end{equation}
where polynomial sensitivity models of degree 0--4 were compared.  The adopted quadratic sensitivity model was
\begin{equation}
    S(x,y) = c_0 + c_x\Delta x + c_y\Delta y
    + c_{xx}\Delta x^2 + c_{xy}\Delta x\Delta y
    + c_{yy}\Delta y^2,
\end{equation}
where $\Delta x=x-x_{\rm ref}$ and $\Delta y=y-y_{\rm ref}$ are the centroid offsets from the reference position. The quadratic model was selected using the Bayesian information criterion (BIC). No additional time-dependent baseline was required. The photometry and corresponding uncertainties were divided by the fitted sensitivity model, and the uncertainties were subsequently rescaled by a common factor based on the reduced $\chi^2$ of a transit fit to the corrected light curve before use in the global analysis. The best-fitting transit model and resulting light curve is shown in Figure \ref{fig:spitzer}.

For the 8 $\mu$m eclipse, aperture photometry was performed directly on the individual BCD images for both HD 80606 and its nearby companion HD 80607. The stellar positions were measured independently in each image using a local background-subtracted centroid, and a 4-pixel aperture was adopted, with the background estimated from an 8-12 pixel annulus. HD 80607, which was observed simultaneously, was used as an independent tracer of instrumental variations during the same observing sequence. Its light curve was best described by a logarithmic time-dependent ramp \citep{Charbonneau2005,Harrington2007,Knutson2007},
\begin{equation}
    R(t)=c_0+c_1\ln(t+\tau),
\end{equation}
together with linear decorrelation terms in detector position and background. The resulting fractional ramp shape, normalised at the eclipse epoch, is denoted $R_{607}(t)$ below.

The rapidly changing thermal emission from HD 80606 b around periastron was represented empirically by a local polynomial. For the adopted cubic model,
\begin{equation}
P(t)
=
d+p_1\Delta t+p_2\Delta t^2+p_3\Delta t^3,
\qquad
\Delta t=t-t_{\rm sec},
\end{equation}
where $d=P(t_{\rm sec})$ is the planet-to-star flux ratio at the eclipse epoch. We define $V(t)$ as the fraction of the planet visible during the eclipse, such that $V=1$ outside eclipse and $V=0$ during full eclipse, with ingress and egress calculated using \texttt{batman} \citep{batman:2015}. The astrophysical component was therefore represented as
$F_{\rm astro}(t)=1+V(t)P(t)$.

We initially fitted the HD 80606 photometry with the instrumental and astrophysical components simultaneously. However, the logarithmic ramp and polynomial planetary-emission baseline were found to be degenerate. We therefore adopted the fractional ramp shape $R_{607}(t)$ measured independently from HD 80607 over the same interval. The final model fitted to HD 80606 was
\begin{equation}
\begin{aligned}
F_{\rm model}(t) ={}&
C\,R_{607}(t)
\left[1+a_x\Delta x+a_y\Delta y+a_B\Delta B\right] \\
&\times \left[1+V(t)P(t)\right].
\end{aligned}
\end{equation}
where $\Delta x=x-x_{\rm ref}$, $\Delta y=y-y_{\rm ref}$, and $\Delta B=B-B_{\rm ref}$ are offsets from the median centroid positions and background level. The normalisation $C$, decorrelation coefficients $a_x$, $a_y$, and $a_B$, planetary-emission coefficients, eclipse depth, and eclipse midpoint remained free parameters. Different polynomial orders and fitting windows were tested, with a cubic model over a $\pm6$ h window adopted based on the BIC and the stability of the recovered eclipse parameters.

The final light curve was detrended using the fitted instrumental model and planetary-emission baseline. The instrumental systematics and time-variable component of the planetary emission were removed while
preserving the secondary eclipse at a constant depth. The resulting unbinned light curve was used in the subsequent \exofast global fit, and is shown in Figure \ref{fig:spitzer}.

\section{JWST photometry}
\label{app:jwst}

We used the publicly available, pipeline-processed JWST white-light curves described in Section~\ref{sec:jwst}, retrieved from the Mikulski Archive for Space Telescopes (MAST) as \texttt{\_whtlt.ecsv} products. These products were generated by the \texttt{white\_light} step of the \texttt{calwebb\_tso3} pipeline, which integrates the extracted time-series spectra over wavelength for each integration. The NIRSpec observation provided separate light curves for the NRS1 and NRS2 detectors, while the MIRI observation provided a single white-light curve. We applied additional processing to remove residual instrumental systematics and slowly varying trends before using the detrended photometry in the global \exofast analysis.

We first rejected outliers from each light curve using a $5\sigma$ median absolute deviation (MAD) criterion. We then visually identified the out-of-eclipse regions on either side of each secondary eclipse and fitted polynomial baselines to these regions. The polynomial orders were selected primarily using the Bayesian information criterion (BIC), while also requiring the baselines to describe the local variations around the eclipses without introducing artificial structure. We adopted fifth- and third-order polynomials for NIRSpec NRS1 and NRS2, respectively, and a fourth-order polynomial for MIRI.

The fitted polynomial baselines were divided out of the unbinned light curves. Each light curve was then normalised so that the stellar flux during the secondary eclipse was unity. We estimated the photometric uncertainties from the robust scatter of the out-of-eclipse residuals. The resulting detrended, unbinned light curves were supplied to the global \exofast fit; the polynomial baseline parameters were not refitted as part of that analysis. Figures~\ref{fig:jwst_nirspec} and \ref{fig:jwst_miri} show the original light curves, adopted baseline models, detrended photometry, and best-fitting eclipse models. The plotted data are binned for visualisation purposes only.
 
\section{General-relativistic timing including adiabatic tides and rotational effects}
\label{app:gr}

In this section we determine the effects on the transit and eclipse timings due to the Newtonian tidal interaction between the planet and the star, and the quadrupole moments of the star and the planet induced by their rotation. Next we compute the effect of the general relativistic (GR) correction, including the well-known orbital precession, at the lowest post-Newtonian (1PN) order. 

To this end we adopt, generalizing Eq.~(16) of~\citet{blanchet2019detecting} (BHL), a quasi-Keplerian parametrization of the orbit. Using polar coordinates $(r,\varphi)$ in the orbital plane,\footnote{In absence of spins or with spins aligned with the orbital angular momentum the motion is planar.} where $\varphi$ is the true anomaly, and a parametrization by the eccentric anomaly $\psi$, we have
\begin{subequations}\label{QKepler}
\begin{align}
	r &= a_r \bigl( 1 - e_r \cos\psi \bigr) ,\\
	n(t-t_\star) &= \psi - e_t \sin\psi \nonumber\\
	& + f_0(v-\psi) + f_1 \sin v + f_2 \sin 2v,\label{keplereq}\\\frac{\varphi-\omega_\star}{K} &= v + g_2 \sin 2v + g_3 \sin 3v + g_4 \sin 4v,
\end{align}
\end{subequations}
where $t_\star$ and $\omega_\star$ are two integration constants,\footnote{In this Appendix our convention for the periastron angle $\omega_\star$ differs from the definition in Table \ref{tab:planet_parameters}, $\omega_\star = \omega_\star^{\text{Table \ref{tab:planet_parameters}}} + \frac{\pi}{2}$.}
and we pose 
\begin{align}\label{vdef}
	v = \psi + 2 \arctan \biggl(\frac{\beta\sin\psi}{1-\beta\cos\psi}\biggr),
\end{align}
with $\beta = [1-(1-e_\varphi^2)^{1/2}]/e_\varphi$. The orbital motion is parametrized by a number of constants: the radial frequency or mean motion $n=2\pi/P$ (where $P$ is the orbital period), the orbital precession $K$, a particular definition of semi-major axis $a_r$, three types of eccentricities $e_r$, $e_t$ and $e_\varphi$, and six additional parameters $f_i$ ($i=0,1,2$) and $g_i$ ($i=2,3,4$). All these constants are related to the orbit's energy $E$ and angular momentum $J$ (projected perpendicularly to the orbital plane) and to the bodies' properties. Rather than $E$ and $J$ we adopt ``Newtonian'' definitions of the semi-major axis $a$ and eccentricity $e$: 
\begin{align}\label{defae}
	a \equiv - \frac{G M\mu}{2 E},\qquad e \equiv \sqrt{1 + \frac{2E h^2}{\mu^3}},
\end{align}
where $h=\frac{J}{G M}$, the total mass is $M=M_\star+M_p$ and the reduced mass is $\mu=M_\star M_p/M$.

The deviations from the Newtonian model with two point masses due to the internal structure (both tidal and rotational effects) and the GR effect are encoded in the expressions of the previous parameters. We have
\begin{subequations}\label{defparameters}
	\begin{align}
		n &= \sqrt{\tfrac{G M}{a^3}}\Bigl(1+\zeta^\text{tide} + \zeta^\text{GR}\Bigr),\\ 
		K &= 1 + k^\text{tide} + k^\text{rot} + k^\text{GR} ,\\
		a_r &= a\Bigl(1 + \xi^\text{tide} + \xi^\text{rot} + \xi^\text{GR}\Bigr),\\
		e_r &= e + \varepsilon_r^\text{tide} + \varepsilon_r^\text{rot} + \varepsilon_r^\text{GR} ,\\ e_t &= e + \varepsilon_t^\text{tide} + \varepsilon_t^\text{rot} + \varepsilon_t^\text{GR} ,\\ 	
		e_\varphi &= e + \varepsilon_\varphi^\text{tide} + \varepsilon_\varphi^\text{rot} + \varepsilon_\varphi^\text{GR} ,\\
		f_i &= f_i^\text{tide} \quad\text{($i=0,1,2$)},\\ 
		g_i &= g_i^\text{tide} \quad\text{($i=2,3,4$)}.
	\end{align}
\end{subequations}
Any other contribution not indicated here is zero. 

The adiabatic tidal effects at the dominant quadrupole level depend on a particular combination (``plus'') of the tidal polarizability parameters defined as (see, \textit{e.g.}, \citealt{ragozzine2009probing})
\begin{align}\label{mu+2}
	\mu_+^{(2)} \equiv \frac{1}{3}\biggl[ \frac{M_\star}{M_p} \,k_p^{(2)} R_p^5 + \frac{M_p}{M_\star} \,k_\star^{(2)} R_\star^5\biggr],
\end{align}
where $k_p^{(2)}$ and $k_\star^{(2)}$ denote the quadrupolar Love numbers of the planet and the star, and $R_p$ and $R_\star$ are the radius. The tidal corrections in all the previous parameters read~\citep{henry2026tides}
\begin{subequations}\label{parameterstide}
	\begin{align}
		\zeta^\mathrm{tide}
		&= - \frac{9(4+e^2)}{2a^5(1-e^2)^{7/2}}
		\,\mu_+^{(2)},\\
		k^\mathrm{tide}
		&= \frac{45(8+12e^2+e^4)}{8a^5(1-e^2)^5}
		\,\mu_+^{(2)},\label{ktide}\\
		\xi^\mathrm{tide}
		&= -\frac{12(1+e^2)}{a^5(1-e^2)^4}
		\,\mu_+^{(2)},\\
		\varepsilon_r^\mathrm{tide}
		&= \frac{3(1+10e^2+5e^4)}{a^5 e(1-e^2)^4}
		\,\mu_+^{(2)},\\
		\varepsilon_t^\mathrm{tide}
		&= \frac{3}{2a^5 e(1-e^2)^{9/2}}
		\notag\\
		&\quad\times\Bigl[
		2\sqrt{1-e^2}
		+12e^2\bigl(-1+\sqrt{1-e^2}\bigr)
		\notag\\
		&\qquad
		+e^4\bigl(9+2\sqrt{1-e^2}\bigr)+3e^6
		\Bigr]\mu_+^{(2)},\\
		\varepsilon_\varphi^\mathrm{tide}
		&= \frac{3(2+60e^2+31e^4)}{2a^5 e(1-e^2)^4}
		\,\mu_+^{(2)},\\
		f_0^\mathrm{tide}
		&= \frac{9(4+e^2)}{2a^5(1-e^2)^{7/2}}
		\,\mu_+^{(2)},\\
		f_1^\mathrm{tide}
		&= \frac{12e}{a^5(1-e^2)^{7/2}}
		\,\mu_+^{(2)},\\
		f_2^\mathrm{tide}
		&= \frac{3e^2}{4a^5(1-e^2)^{7/2}}
		\,\mu_+^{(2)},\\
		g_2^\mathrm{tide}
		&= \frac{3e^2(15+2e^2)}{4a^5(1-e^2)^5}
		\,\mu_+^{(2)},\\
		g_3^\mathrm{tide}
		&= \frac{3e^3}{2a^5(1-e^2)^5}
		\,\mu_+^{(2)},\\
		g_4^\mathrm{tide}
		&= \frac{3e^4}{32a^5(1-e^2)^5}
		\,\mu_+^{(2)}.
	\end{align}
\end{subequations}

Next we give the rotational contributions, which depend on the following combination of parameters for the planet and star (\textit{e.g.}, \citealt{ragozzine2009probing})
\begin{align}\label{lambda+2}
	\lambda_+^{(2)} \equiv k_p^{(2)} \frac{R_p^5\Omega_p^2}{G M_p} + k_\star^{(2)} \frac{R_\star^5\Omega_\star^2}{G M_\star} ,
\end{align}
where the internal structure is again described by the Love numbers and $\Omega_p=2\pi/P_p$ and $\Omega_\star=2\pi/P_\star$ denote the angular rotation frequencies. We have~\citep{henry2023spin}
\begin{subequations}\label{parametersrot}
\begin{align}
	k^\text{rot} &= \frac{1}{2a^2(1-e^2)^2} \lambda_+^{(2)},\label{krot}\\
     \xi^\text{rot} &= - \frac{1}{6a^2(1-e^2)} \lambda_+^{(2)},\\
	\varepsilon_r^\text{rot} &= \frac{1+e^2}{6a^2e(1-e^2)} \lambda_+^{(2)},\\ 
	\varepsilon_t^\text{rot} &= \frac{1}{6a^2e(1-e^2)} \lambda_+^{(2)},\\ 	
	\varepsilon_\varphi^\text{rot} &= \frac{1+2e^2}{6a^2e(1-e^2)} \lambda_+^{(2)}.
\end{align}
\end{subequations}

Finally we recall from BHL the corrections due to the GR effect at the leading 1PN level:
\begin{subequations}\label{parametersGR}
\begin{align}
	\zeta^\text{GR} &= \frac{G M}{8a c^2}\bigl(-15+\nu\bigr),\label{zeta}\\
	k^\text{GR} &= \frac{3G M}{a c^2(1-e^2)},\label{kGR}\\
		\xi^\text{GR} &= \frac{G M}{4a c^2}\bigl(-7+\nu\bigr),\\
	\varepsilon_r^\text{GR} &= \frac{G M}{8a c^2}\biggl[\frac{9+\nu}{e}+\bigl(15-5\nu\bigr)e\biggr] ,\\
	\varepsilon_t^\text{GR} &= \frac{G M}{8a c^2}\biggl[\frac{9+\nu}{e}+\bigl(-17+7\nu\bigr)e\biggr],\\	\varepsilon_\varphi^\text{GR} &= \frac{G M}{8a c^2}\biggl[\frac{9+\nu}{e}+\bigl(15-\nu\bigr)e\biggr] ,
\end{align}
\end{subequations}
where $\nu=\mu/M$ denotes the symmetric mass ratio.

We can compare the orders of magnitude of these effects in the apsidal precession:
\begin{align}\label{apsidal}
	\dot{\omega} \equiv n \,k = \dot{\omega}^\text{GR} + \dot{\omega}^\text{tide} + \dot{\omega}^\text{rot},
\end{align}
where $n=\sqrt{G M/a^3}$ at the leading order. Using the orbital elements in Table~\ref{tab:planet_parameters}, with the Love numbers $k_p^{(2)}\simeq 0.6$ and $k_\star^{(2)}\simeq 0.1$ and the rotation frequencies $\Omega_p\simeq 2\pi/93\,h^{-1}$ and $\Omega_\star\simeq 2\pi/40\,d^{-1}$, we find
\begin{subequations}
\begin{align}\label{apsidal2}
	\dot{\omega}^\text{GR} &\simeq 210\,\mathrm{arcsec~century^{-1}},\\
    \dot{\omega}^\text{tide} &\simeq 97\,\mathrm{arcsec~century^{-1}},\\
    \dot{\omega}^\text{rot} &\simeq 3.5\,\mathrm{arcsec~century^{-1}},
\end{align}
\end{subequations}
so that in the case of HD 80606 b the GR precession dominates over the tidal and rotational precessions.

Having the above explicit solution of the motion we compute all the corrections to the timing of transits and eclipses. We follow and improve the steps in BHL. The impact parameter of the planet is the minimal distance between the trajectory of the planet projected perpendicularly to the line of sight and the center of the star. It reads
\begin{subequations}\label{b}
	\begin{align}
		&b = r(\pi)\sin\Omega \sin i && \text{(for transits)},\\
		&\bar{b} = r(0)\sin\Omega \sin i &&\text{(for eclipses)},
	\end{align}
\end{subequations}
where the overbar distinguishes the eclipse from the transit, and where $r(\pi)$ and $r(0)$ are the values of the radial coordinate of the planet when the polar angle (or true anomaly) takes the values $\varphi=\pi$ and $\varphi=0$ [modulo $2\pi$], when the transit and eclipse occur. We thus require the function $r(\varphi)$ which is obtained as direct consequence of Eqs.~\eqref{QKepler}:
\begin{subequations}\label{rphi}
		\begin{align}
			r(\varphi) &= a_r\left[1-e_r
			\frac{e_\varphi+\cos \Phi}{1+e_\varphi\cos \Phi}\right],\\
			\Phi &\equiv \frac{\phi}{K}-g_2\sin2\phi
			-g_3\sin3\phi-g_4\sin4\phi,
		\end{align}
\end{subequations}
where we pose $\phi=\varphi-\omega_\star$. 

In the Newtonian model with no internal structure (zero Love numbers), the impact parameters reduce to
\begin{subequations}\label{b0}
	\begin{align}
		&b_0 = \frac{a(1-e^2)}{1-e \cos\omega_\star}\sin\Omega \sin i&& \text{(for transits)},\\
		&\bar{b}_0 = \frac{a(1-e^2)}{1+e \cos\omega_\star}\sin\Omega \sin i &&\text{(for eclipses)},
	\end{align}
\end{subequations}
where $a$ and $e$ are defined by~\eqref{defae}. With the internal structure and the GR corrections included, the impact parameters defined by~\eqref{b} are modified as
\begin{equation}\label{deltab}
	b = b_0 + \delta b, \qquad \bar{b} = \bar{b}_0 + \delta \bar{b},
\end{equation}
and the corrections $\delta b$ and $\delta \bar{b}$ can straightforwardly be computed from $r(\varphi)$ given in~\eqref{rphi} by inserting all the parameters with their explicit expressions~\eqref{parameterstide},~\eqref{parametersrot} and~\eqref{parametersGR} and performing a linear perturbation.

The transits and eclipses correspond to particular values of the vertical coordinate $Y$ in the particular coordinate system $\{X, Y\}$ defined in BHL (see Fig.~1 there), where $X$ and $Y$ lie in the orbital plane with $X$ in the observer's direction. In the polar coordinates we have
\begin{equation}\label{Ydef}
	Y \equiv r\sin\varphi.
\end{equation}
The positions of the planet $\{T_1 , T_2 , T_m, T_3 , T_4\}$ on its trajectory are respectively defined to be the entry of the transit, the start of the full transit, the mid-term point of closest approach to the star's center, and the exits of the full and partial transits. Similarly for the points $\{\bar{T}_1 , \bar{T}_2 , \bar{T}_m, \bar{T}_3 , \bar{T}_4\}$ which correspond to the eclipse, see Fig.~1 in BHL. The conditions on the coordinate $Y$ at the passage at these points read, for transits,
\begin{equation}\label{Yconditionstransit}
    Y = \left\{\begin{array}{lll} \frac{M_\star}{M}\biggl[\sqrt{(R_\star+R_p)^2-b^2} + b\,\cot i\biggr] &&\hspace{-0.6cm}\text{($T_1$)},\\[0.4cm] \frac{M_\star}{M}\biggl[\sqrt{(R_\star-R_p)^2-b^2} + b\,\cot i\biggr] &&\hspace{-0.6cm}\text{($T_2$)},\\[0.4cm] 
	b\,\cot i 
	&&\hspace{-0.6cm}\text{($T_m$)},\\[0.4cm] \frac{M_\star}{M}\biggl[-\sqrt{(R_\star-R_p)^2-b^2} + b\,\cot i\biggr] &&\hspace{-0.6cm}\text{($T_3$)},\\[0.4cm] \frac{M_\star}{M}\biggl[-\sqrt{(R_\star+R_p)^2-b^2} + b\,\cot i\biggr] &&\hspace{-0.6cm}\text{($T_4$)},
	\end{array}\right.
\end{equation}
and, for eclipses,
\begin{equation}\label{Yconditionseclipse}
	Y = \left\{\begin{array}{lll} \frac{M_\star}{M}\biggl[\sqrt{(R_\star+R_p)^2-\bar{b}^2} + \bar{b}\,\cot i\biggr] &&\hspace{-0.6cm}\text{($\bar{T}_4$)},\\[0.4cm] \frac{M_\star}{M}\biggl[\sqrt{(R_\star-R_p)^2-\bar{b}^2} + \bar{b}\,\cot i\biggr] &&\hspace{-0.6cm}\text{($\bar{T}_3$)},\\[0.4cm] 
	\bar{b}\,\cot i 
	&&\hspace{-0.6cm}\text{($\bar{T}_m$)},\\[0.4cm] \frac{M_\star}{M}\biggl[-\sqrt{(R_\star-R_p)^2-\bar{b}^2} + \bar{b}\,\cot i\biggr] &&\hspace{-0.6cm}\text{($\bar{T}_2$)},\\[0.4cm] \frac{M_\star}{M}\biggl[-\sqrt{(R_\star+R_p)^2-\bar{b}^2} + \bar{b}\,\cot i\biggr] &&\hspace{-0.6cm}\text{($\bar{T}_1$)}.
	\end{array}\right.
\end{equation}
The coordinate $Y$ defined by~\eqref{Ydef} can be obtained in closed-form as a function of the eccentric anomaly $\psi$ from the solution of the motion~\eqref{QKepler}. We get
\begin{align}\label{Y}
	Y(\psi) &= a_r\Biggl\{\frac{1-e_r\cos\psi}{1-e_\phi\cos\psi}\\
	&\times\Bigl[\sqrt{1-e_\phi^2}\sin\psi\cos\omega_\star + (\cos\psi-e_\phi)\sin\omega_\star\Bigr] \nonumber\\
	& \qquad + \bigl(k \,v + g_2 \sin 2v + g_3 \sin 3v + g_4 \sin 4v\bigr)\nonumber\\
	&\times\Bigl[(\cos\psi-e_\phi)\cos\omega_\star - \sqrt{1-e_\phi^2}\sin\psi\sin\omega_\star\Bigr]\Biggr\}.\nonumber
\end{align}
For the reference Newtonian solution we have
\begin{equation}\label{Y0}
Y_0(\psi) = a \left[\sqrt{1-e^2}\sin\psi\cos\omega_\star+(\cos\psi-e)\sin\omega_\star\right].
\end{equation}
With~\eqref{Y}--\eqref{Y0} we define the correction $\delta Y$ such that
\begin{equation}\label{Y0prime}
	Y(\psi) = Y_0(\psi) + \delta Y(\psi),
\end{equation}
which is easy to obtain in closed-form using the explicit expressions of the parameters~\eqref{parameterstide},~\eqref{parametersrot} and~\eqref{parametersGR}. 

The perturbation calculation goes as follows. For simplicity we restrict ourself to the calculation of the mid-transit points $T_m$, for which the equation to solve is $Y(\psi) = b \cot i$, as we see from~\eqref{Yconditionstransit}, where we recall that the impact parameter has been computed before, in~\eqref{deltab}. We first solve the Newtonian problem by determining $\psi_0$ such that $Y_0(\psi_0)=b_0\cot i$ where $b_0$ is the Newtonian impact parameter in~\eqref{b0} and $Y_0$ is given by~\eqref{Y0}. Then we look for the perturbation of the eccentric anomaly as $\psi = \psi_0 + \delta\psi$ and obtain
\begin{equation}\label{deltapsi}
	\delta\psi = \frac{\delta b \cot i -\delta Y(\psi_0)}{Y'_0(\psi_0)},
\end{equation}
where $\delta b$ is the perturbation in the impact parameter in~\eqref{deltab} and $Y'_0(\psi)\equiv\dd Y_0/\dd\psi$ is easily obtained from~\eqref{Y0}. For the other points the procedure is the same with just the replacement of $\delta b \cot i$ by the appropriate expression computed from the right-hand sides of~\eqref{Yconditionstransit} or~\eqref{Yconditionseclipse}. Finally, knowing the perturbation $\delta\psi$, it remains to compute the modification of the instants of transit and eclipses using the generalized Kepler equation~\eqref{keplereq}. To Newtonian order we have
\begin{equation}
	t_0 = t_\star + \sqrt{\frac{a^3}{G M}}\Bigl[\psi_0 - e \sin \psi_0\Bigr],
\end{equation}
while to first order $t=t_0+\delta t$ with
\begin{align}\label{deltat}
	\delta t &= \sqrt{\frac{a^3}{G M}}\Bigl[ \bigl(1-e_t\cos\psi_0\bigr)\delta\psi - \bigl(1-e_t\sin\psi_0\bigr)\zeta \nonumber\\
	&\qquad + f_0(v_0-\psi_0) + f_1 \sin v_0 + f_2 \sin 2v_0 \Bigr].
\end{align}
Here $v_0$ is given by~\eqref{vdef} but computed with $\psi_0$. The calculation is done for a particular orbit but is trivially generalized to an arbitrary orbit $N$. By repeating the calculation for all the points described in~\eqref{Yconditionstransit}--\eqref{Yconditionseclipse} one finally obtains the relativistic time shifts for transits 
\begin{subequations}\label{indiviualshifts}
\begin{align}
	&\bigl\{\delta t_1(N), \delta t_2(N),\delta t_m(N), \delta t_3(N),\delta t_4(N)\bigr\},
\end{align}
and for eclipses 
\begin{align}
	&\bigl\{\delta \bar{t}_1(N), \delta \bar{t}_2(N),\delta \bar{t}_m(N), \delta \bar{t}_3(N),\delta \bar{t}_4(N)\bigr\}.
\end{align}
\end{subequations}
It is important to note that the individual timing shifts $\delta t_i$ and $\delta \bar{t}_i$ represent the differences between the real model and a Newtonian model, and as such are not directly observable because they depend on the reference Newtonian model. To obtain a directly measurable observable, we shall consider the evolution of the time interval between a transit and an eclipse, given by a combination of the individual shifts \eqref{indiviualshifts} which is independent of the reference model. This is detailed in Sec.~\ref{sec:theoretical predictions}.

\section{\exofast global Newtonian fit  results}

In this appendix, we provide the complete set of parameters obtained from the global Newtonian \exofast fit described in Section 3. The results are reported as posterior median values with their corresponding 68\% credible intervals. Table E.1 presents the stellar parameters of HD 80606, while Table E.2 gives the wavelength-dependent parameters, including limb-darkening coefficients and eclipse depths. Table E.3 lists the instrument-dependent radial-velocity offsets and jitter terms, and Table E.4 summarises the photometric parameters associated with the individual transit and eclipse light curves. Finally, Table E.5 presents the planetary and orbital parameters of HD 80606 b.

\label{app:exofast}

\begin{table}
\caption{Median values and 68\% confidence intervals for fitted stellar parameters of HD 80606.}
\centering
\small
\setlength{\tabcolsep}{3pt}
\begin{tabular}{lcc}
\hline\hline
Parameter & Description & Values\\
\hline
~~~~$M_*$\dotfill &Mass (\msun)\dotfill &$1.009^{+0.047}_{-0.045}$\\
~~~~$R_*$\dotfill &Radius (\rsun)\dotfill &$1.049\pm0.016$\\
~~~~$R_{\rm *,SED}$\dotfill &Radius$^{1}$ (\rsun)\dotfill &$1.061^{+0.010}_{-0.011}$\\
~~~~$L_*$\dotfill &Luminosity (\lsun)\dotfill &$0.936^{+0.027}_{-0.026}$\\
~~~~$F_{Bol}$\dotfill &Bolometric Flux (cgs)\dotfill &$6.89^{+0.20}_{-0.19} \times 10^{-9}$\\
~~~~$\rho_*$\dotfill &Density (cgs)\dotfill &$1.231^{+0.021}_{-0.020}$\\
~~~~$\log{g}$\dotfill &Surface gravity (cgs)\dotfill &$4.4001\pm0.0086$\\
~~~~$T_{\rm eff}$\dotfill &Effective temperature (K)\dotfill &$5541^{+51}_{-50}$\\
~~~~$T_{\rm eff,SED}$\dotfill &Effective temperature$^{1}$ (K)\dotfill &$5515^{+36}_{-34}$\\
~~~~$[{\rm Fe/H}]$\dotfill &Metallicity (dex)\dotfill &$0.352^{+0.074}_{-0.076}$\\
~~~~$[{\rm Fe/H}]_{0}$\dotfill &Initial Metallicity$^{2}$ \dotfill &$0.345^{+0.069}_{-0.071}$\\
~~~~$Age$\dotfill &Age (Gyr)\dotfill &$6.9^{+2.7}_{-2.3}$\\
~~~~$EEP$\dotfill &Equal Evolutionary Phase$^{3}$ \dotfill &$386^{+18}_{-26}$\\
~~~~$A_V$\dotfill &V-band extinction (mag)\dotfill &$0.022^{+0.014}_{-0.015}$\\
~~~~$\sigma_{\rm SED}$\dotfill &SED photometry error scaling \dotfill &$1.32^{+0.47}_{-0.30}$\\
~~~~$\varpi$\dotfill &Parallax (mas)\dotfill &$15.174\pm0.020$\\
~~~~$d$\dotfill &Distance (pc)\dotfill &$65.900^{+0.086}_{-0.085}$\\
\hline
\label{tab:exofast_stellar}
\end{tabular}
\tablefoot{See Table 3 in \citet{Eastman:2019} for a detailed description of all parameters. (1) This value ignores the systematic error and is for reference only. (2) The metallicity of the star at birth. (3) Corresponds to static points in a star's evolutionary history. See Section 2 in \citet{Dotter:2016}.}
\end{table}


\begin{table}
\caption{Wavelength-dependent parameters.}
\label{tab:wavelength_parameters}
\centering
\small
\setlength{\tabcolsep}{3pt}
\begin{tabular}{@{}lcccc@{}}
\hline\hline
Bandpass
    & $u_1$
    & $u_2$
    & $A_T$
    & $\delta_S$ \\
    & 
    & 
    & (ppm)
    & (ppm) \\
\hline
$R$
    & $0.461\pm0.046$
    & $0.261\pm0.048$
    & --
    & -- \\

$4.5\,\mu\mathrm{m}$
    & $0.031^{+0.032}_{-0.022}$
    & $0.083^{+0.043}_{-0.042}$
    & --
    & -- \\

$8.0\,\mu\mathrm{m}$
    & $0.044^{+0.032}_{-0.027}$
    & $0.102\pm0.035$
    & $1239\pm17$
    & $1239\pm17$ \\

TESS
    & $0.355^{+0.041}_{-0.042}$
    & $0.251\pm0.042$
    & --
    & -- \\
\hline
\end{tabular}
\tablefoot{%
$u_1$ and $u_2$ are the linear and quadratic limb-darkening
coefficients, respectively. $A_T$ is the thermal emission from the
planet, and $\delta_S$ is the measured eclipse depth.
}
\end{table}


\begin{table}
\caption{Telescope-dependent radial-velocity parameters.}
\label{tab:telescope_parameters}
\centering
\small
\begin{tabular}{@{}lccc@{}}
\hline\hline
Instrument
    & $\gamma_{\rm rel}$
    & $\sigma_J$
    & $\sigma_J^2$ \\
    & ($\mathrm{m\,s^{-1}}$)
    & ($\mathrm{m\,s^{-1}}$)
    & ($\mathrm{m^2\,s^{-2}}$) \\
\hline
APF
    & $147.8^{+4.5}_{-4.8}$
    & $8.7^{+9.6}_{-3.9}$
    & $75^{+260}_{-52}$ \\

ELODIE
    & $3777.5\pm2.3$
    & $14.2^{+2.4}_{-2.2}$
    & $200^{+73}_{-58}$ \\

HARPS-N
    & $3934.7^{+2.5}_{-2.2}$
    & $4.4^{+3.8}_{-2.1}$
    & $18^{+47}_{-14}$ \\

HIRES$_j$
    & $-76.91^{+0.49}_{-0.50}$
    & $3.68^{+0.37}_{-0.32}$
    & $13.5^{+2.8}_{-2.3}$ \\

HIRES$_k$
    & $-80.9\pm1.1$
    & $6.41^{+0.99}_{-0.83}$
    & $41.1^{+14}_{-9.9}$ \\

HRS
    & $-31.9^{+4.1}_{-3.9}$
    & $17.7^{+3.5}_{-2.8}$
    & $312^{+140}_{-90}$ \\

SOPHIE
    & $3905.1^{+1.1}_{-1.2}$
    & $5.39^{+1.0}_{-0.92}$
    & $29.0^{+12}_{-9.0}$ \\
\hline
\end{tabular}
\tablefoot{%
$\gamma_{\rm rel}$ is the relative radial-velocity offset,
$\sigma_J$ is the radial-velocity jitter, and $\sigma_J^2$ is the
radial-velocity jitter variance.
}
\end{table}

\begin{table}
\caption{Transit and eclipse light-curve parameters.}
\label{tab:transit_parameters}
\centering
\small
\setlength{\tabcolsep}{4pt}
\begin{tabular}{@{}lcc@{}}
\hline\hline
Dataset & $\sigma^2$ & $F_0$ \\
\hline
Spitzer 8$\mu$m (2007)
    & $8.8^{+1.5}_{-1.4}\times10^{-7}$
    & $0.999998\pm0.000045$ \\
OHP (2009)
    & $6.5^{+4.5}_{-4.1}\times10^{-7}$
    & $1.00058\pm0.00013$ \\
Spitzer 4.5$\mu$m (2010)
    & $0.3^{+1.2}_{-1.1}\times10^{-7}$
    & $1.000055\pm0.000066$ \\
TESS (2020)
    & $(1.69\pm0.11)\times10^{-7}$
    & $0.999985\pm0.000018$ \\
MIRI (2025)
    & $2.92^{+0.56}_{-0.55}\times10^{-8}$
    & $1.000000\pm0.000016$ \\
\hline
\end{tabular}
\tablefoot{
$\sigma^2$ is the additional flux variance and
$F_0$ is the baseline normalised flux.
For the TESS light curve, the fitted dilution factor
is $A_D = 0.4608^{+0.0084}_{-0.0086}$,
where $A_D = F_2/(F_1+F_2)$ and $F_1$ and $F_2$
are the fluxes of the target and contaminating
star, respectively.
}
\end{table}

\begin{table*}
\caption{Median values and 68\% confidence intervals for fitted planetary parameters of HD 80606 b.}
\centering
\begin{tabular}{lcc}
\hline\hline
Parameter & Description & Values\\
\hline
~~~~$P$\dotfill &Period (days)\dotfill &$111.437252\pm0.000022$\\
~~~~$R_P$\dotfill &Radius (\rj)\dotfill &$1.011^{+0.017}_{-0.016}$\\
~~~~$M_P$\dotfill &Mass (\mj)\dotfill &$4.09^{+0.13}_{-0.12}$\\
~~~~$T_C$\dotfill &Observed Time of conjunction$^{1}$ (\bjdtdb)\dotfill &$2458888.07964^{+0.00079}_{-0.00080}$\\
~~~~$T_C$\dotfill &Model Time of conjunction$^{1,2}$ (\tjdtdb)\dotfill &$2458888.07794\pm0.00080$\\
~~~~$T_T$\dotfill &Model time of min proj sep$^{2,3}$ (\tjdtdb)\dotfill &$2456770.76135\pm0.00066$\\
~~~~$T_0$\dotfill &Obs time of min proj sep$^{3,4}$ (\bjdtdb)\dotfill &$2456770.76305\pm0.00066$\\
~~~~$a$\dotfill &Semi-major axis (AU)\dotfill &$0.4551^{+0.0069}_{-0.0068}$\\
~~~~$i$\dotfill &Inclination (Degrees)\dotfill &$89.2289^{+0.0071}_{-0.0070}$\\
~~~~$e$\dotfill &Eccentricity \dotfill &$0.93116\pm0.00030$\\
~~~~$\omega_*$\dotfill &Arg of periastron (Degrees)\dotfill &$-58.668\pm0.085$\\
~~~~$\dot{\omega}_{\rm GR}$\dotfill &Computed GR precession ($^\circ$/century)\dotfill &$0.0583^{+0.0018}_{-0.0017}$\\
~~~~$T_{\rm eq}$\dotfill &Equilibrium temp$^{5}$ (K)\dotfill &$405.7\pm3.7$\\
~~~~$\tau_{\rm circ}$\dotfill &Tidal circ timescale (Gyr)\dotfill &$2.87\pm0.16$\\
~~~~$K$\dotfill &RV semi-amplitude (m/s)\dotfill &$469.3\pm1.2$\\
~~~~$R_P/R_*$\dotfill &Radius of planet in stellar radii \dotfill &$0.09902^{+0.00052}_{-0.00051}$\\
~~~~$a/R_*$\dotfill &Semi-major axis in stellar radii \dotfill &$93.27\pm0.52$\\
~~~~$\delta$\dotfill &$\left(R_P/R_*\right)^2$ \dotfill &$0.00981\pm0.00010$\\
~~~~$\delta_{\rm R}$\dotfill &Transit depth in R (frac)\dotfill &$0.00910\pm0.00013$\\
~~~~$\delta_{\rm 4.5\mu m}$\dotfill &Transit depth in $4.5\mu m$ (frac)\dotfill &$0.009763^{+0.000097}_{-0.000096}$\\
~~~~$\delta_{\rm 8.0\mu m}$\dotfill &Transit depth in $8.0\mu m$ (frac)\dotfill &$0.00975\pm0.00011$\\
~~~~$\delta_{\rm TESS}$\dotfill &Transit depth in TESS (frac)\dotfill &$0.00930\pm0.00012$\\
~~~~$\tau$\dotfill &In/egress transit duration (days)\dotfill &$0.1199\pm0.0021$\\
~~~~$T_{14}$\dotfill &Total transit duration (days)\dotfill &$0.4993\pm0.0022$\\
~~~~$T_{FWHM}$\dotfill &FWHM transit duration (days)\dotfill &$0.3794^{+0.0025}_{-0.0024}$\\
~~~~$b$\dotfill &Transit impact parameter \dotfill &$0.8155^{+0.0036}_{-0.0037}$\\
~~~~$b_S$\dotfill &Eclipse impact parameter \dotfill &$0.09294\pm0.00045$\\
~~~~$\tau_S$\dotfill &In/egress eclipse duration (days)\dotfill &$0.007682^{+0.000059}_{-0.000058}$\\
~~~~$T_{S,14}$\dotfill &Total eclipse duration (days)\dotfill &$0.08458^{+0.00046}_{-0.00045}$\\
~~~~$T_{S,FWHM}$\dotfill &FWHM eclipse duration (days)\dotfill &$0.07690\pm0.00041$\\
~~~~$\delta_{S,2.5\mu m}$\dotfill &BB eclipse depth at 2.5$\mu$m (ppm)\dotfill &$0.0124^{+0.0015}_{-0.0013}$\\
~~~~$\delta_{S,5.0\mu m}$\dotfill &BB eclipse depth at 5.0$\mu$m (ppm)\dotfill &$5.56^{+0.31}_{-0.29}$\\
~~~~$\delta_{S,7.5\mu m}$\dotfill &BB eclipse depth at 7.5$\mu$m (ppm)\dotfill &$36.2^{+1.3}_{-1.2}$\\
~~~~$\rho_P$\dotfill &Density (cgs)\dotfill &$4.90^{+0.14}_{-0.13}$\\
~~~~$logg_P$\dotfill &Surface gravity (cgs)\dotfill &$3.9960^{+0.0070}_{-0.0071}$\\
~~~~$\Theta$\dotfill &Safronov Number \dotfill &$3.642\pm0.061$\\
~~~~$\fave$\dotfill &Incident Flux (\fluxcgs)\dotfill &$0.00299\pm0.00011$\\
~~~~$T_S$\dotfill &Observed Time of eclipse$^{1}$ (\bjdtdb)\dotfill &$2458882.22922^{+0.00041}_{-0.00040}$\\
~~~~$T_S$\dotfill &Model Time of eclipse$^{1,2}$ (\tjdtdb)\dotfill &$2458882.22941^{+0.00041}_{-0.00040}$\\
~~~~$T_E$\dotfill &Model time of sec min proj sep$^{2,3}$ (\tjdtdb)\dotfill &$2460999.53722\pm0.00023$\\
~~~~$T_{E,0}$\dotfill &Obs time of sec min proj sep$^{3,4}$ (\bjdtdb)\dotfill &$2460999.53703\pm0.00023$\\
~~~~$T_P$\dotfill &Time of Periastron (\tjdtdb)\dotfill &$2458882.3618^{+0.0014}_{-0.0013}$\\
~~~~$T_A$\dotfill &Time of asc node (\tjdtdb)\dotfill &$2458882.6459\pm0.0026$\\
~~~~$T_D$\dotfill &Time of desc node (\tjdtdb)\dotfill &$2458992.273\pm0.013$\\
~~~~$V_c/V_e$\dotfill &Scaled velocity \dotfill &$1.7817\pm0.0047$\\
~~~~$e\cos{\omega_*}$\dotfill & \dotfill &$0.4842\pm0.0010$\\
~~~~$e\sin{\omega_*}$\dotfill & \dotfill &$-0.79537^{+0.00098}_{-0.00097}$\\
~~~~$M_P\sin i$\dotfill &Minimum mass (\mj)\dotfill &$4.08^{+0.13}_{-0.12}$\\
~~~~$M_P/M_*$\dotfill &Mass ratio \dotfill &$0.003866^{+0.000060}_{-0.000059}$\\
~~~~$d/R_*$\dotfill &Separation at mid transit \dotfill &$60.59\pm0.33$\\
\hline
\label{tab:exofast_planet}
\end{tabular}
\tablefoot{See Table 3 in \citet{Eastman:2019} for a detailed description of all parameters. (1) Time of conjunction is commonly reported as the ``transit time''. (2) \tjdtdb is the target's barycentric frame and corrects for light travel time. (3) Time of minimum projected separation is a more correct ``transit time''. (4) At the epoch that minimizes the covariance between $T_C$ and Period. (5) Assumes no albedo and perfect redistribution.}
\end{table*}

\end{appendix}

\end{document}